
%
\def\SpringerMacroPackageNameATest{AA}%
\let\next\relax
\ifx\SpringerMacroPackageNameA\undefined
  \message{Loading the \SpringerMacroPackageNameATest\space
           macro package from Springer-Verlag...}%
\else
  \ifx\SpringerMacroPackageNameA\SpringerMacroPackageNameATest
    \message{\SpringerMacroPackageNameA\space macro package
             from Springer-Verlag already loaded.}%
    \let\next\endinput
  \else
    \message{DANGER: \SpringerMacroPackageNameA\space from
             Springer-Verlag already loaded, will try to proceed.}%
  \fi
\fi
\next
\def\SpringerMacroPackageNameA{AA}%
\newskip\mathindent      \mathindent=0pt
\newskip\tabefore \tabefore=20dd plus 10pt minus 5pt      
\newskip\taafter  \taafter=10dd                           
\newskip\tbbeforeback    \tbbeforeback=-20dd              
\newskip\tbbefore        \tbbefore=17pt plus 7pt minus3pt 
\newskip\tbafter         \tbafter=8pt                     
\newskip\tcbeforeback    \tcbeforeback=-3pt               
\advance\tcbeforeback by -10dd                            
\newskip\tcbefore        \tcbefore=10dd plus 5pt minus 1pt
\newskip\tcafter         \tcafter=6pt                     
\newskip\tdbeforeback    \tdbeforeback=-3pt                  
\advance\tdbeforeback by -10dd                               
\newskip\tdbefore        \tdbefore=10dd plus 4pt minus 1pt   
\newskip\petitsurround
\petitsurround=6pt\relax
\newskip\ackbefore      \ackbefore=10dd plus 5pt             
\newskip\ackafter       \ackafter=6pt                        
\newdimen\itemindent    \newdimen\itemitemindent
\itemindent=1.5em       \itemitemindent=2\itemindent
 \font \tatt            = cmbx10 scaled \magstep3
 \font \tats            = cmbx10 scaled \magstep1
 \font \tamt            = cmmib10 scaled \magstep3
 \font \tams            = cmmib10 scaled \magstep1
 \font \tamss           = cmmib10
 \font \tast            = cmsy10 scaled \magstep3
 \font \tass            = cmsy10 scaled \magstep1
 \font \tbtt            = cmbx10 scaled \magstep2
 \font \tbmt            = cmmib10 scaled \magstep2
 \font \tbst            = cmsy10 scaled \magstep2
\catcode`@=11    
\vsize=23.5truecm
\hoffset=-1true cm
\voffset=-1true cm
\normallineskip=1dd
\normallineskiplimit=0dd
\newskip\ttglue%
\def\ifundefin@d#1#2{%
\expandafter\ifx\csname#1#2\endcsname\relax}
\def\getf@nt#1#2#3#4{%
\ifundefin@d{#1}{#2}%
\global\expandafter\font\csname#1#2\endcsname=#3#4%
\fi\relax
}
\newfam\sffam
\newfam\scfam
\def\makesize#1#2#3#4#5#6#7{%
 \getf@nt{rm}{#1}{cmr}{#2}%
 \getf@nt{rm}{#3}{cmr}{#4}%
 \getf@nt{rm}{#5}{cmr}{#6}%
 \getf@nt{mi}{#1}{cmmi}{#2}%
 \getf@nt{mi}{#3}{cmmi}{#4}%
 \getf@nt{mi}{#5}{cmmi}{#6}%
 \getf@nt{sy}{#1}{cmsy}{#2}%
 \getf@nt{sy}{#3}{cmsy}{#4}%
 \getf@nt{sy}{#5}{cmsy}{#6}%
 \skewchar\csname mi#1\endcsname ='177
 \skewchar\csname mi#3\endcsname ='177
 \skewchar\csname mi#5\endcsname ='177
 \skewchar\csname sy#1\endcsname ='60
 \skewchar\csname sy#3\endcsname='60
 \skewchar\csname sy#5\endcsname='60
\expandafter\def\csname#1size\endcsname{%
 \normalbaselineskip=#7
 \normalbaselines
 \setbox\strutbox=\hbox{\vrule height0.75\normalbaselineskip%
    depth0.25\normalbaselineskip width0pt}%
 \textfont0=\csname rm#1\endcsname
 \scriptfont0=\csname rm#3\endcsname
 \scriptscriptfont0=\csname rm#5\endcsname
    \def\oldstyle{\fam1\csname mi#1\endcsname}%
 \textfont1=\csname mi#1\endcsname
 \scriptfont1=\csname mi#3\endcsname
 \scriptscriptfont1=\csname mi#5\endcsname
 \textfont2=\csname sy#1\endcsname
 \scriptfont2=\csname sy#3\endcsname
 \scriptscriptfont2=\csname sy#5\endcsname
 \textfont3=\tenex\scriptfont3=\tenex\scriptscriptfont3=\tenex
   \def\rm{%
 \fam0\csname rm#1\endcsname%
   }%
   \def\it{%
 \getf@nt{it}{#1}{cmti}{#2}%
 \textfont\itfam=\csname it#1\endcsname
 \fam\itfam\csname it#1\endcsname
   }%
   \def\sl{%
 \getf@nt{sl}{#1}{cmsl}{#2}%
 \textfont\slfam=\csname sl#1\endcsname
 \fam\slfam\csname sl#1\endcsname}%
   \def\bf{%
 \getf@nt{bf}{#1}{cmbx}{#2}%
 \getf@nt{bf}{#3}{cmbx}{#4}%
 \getf@nt{bf}{#5}{cmbx}{#6}%
 \textfont\bffam=\csname bf#1\endcsname
 \scriptfont\bffam=\csname bf#3\endcsname
 \scriptscriptfont\bffam=\csname bf#5\endcsname
 \fam\bffam\csname bf#1\endcsname}%
   \def\tt{%
 \getf@nt{tt}{#1}{cmtt}{#2}%
 \textfont\ttfam=\csname tt#1\endcsname
 \fam\ttfam\csname tt#1\endcsname
 \ttglue=.5em plus.25em minus.15em
   }%
  \def\sf{%
\getf@nt{sf}{#1}{cmss}{10 at #2pt}%
\textfont\sffam=\csname sf#1\endcsname
\fam\sffam\csname sf#1\endcsname}%
   \def\sc{%
 \getf@nt{sc}{#1}{cmcsc}{10 at #2pt}%
 \textfont\scfam=\csname sc#1\endcsname
 \fam\scfam\csname sc#1\endcsname}%
\rm }}
\makesize{IXf}{9}{VIf}{6}{Vf}{5}{10.00dd}
\def\normalsize{\IXfsize
\def\sf{%
   \getf@nt{sf}{IXf}{cmss}{9}%
   \getf@nt{sf}{VIf}{cmss}{10 at 6pt}%
   \getf@nt{sf}{Vf}{cmss}{10 at 5pt}%
   \textfont\sffam=\csname sfIXf\endcsname
   \scriptfont\sffam=\csname sfVIf\endcsname
   \scriptscriptfont\sffam=\csname sfVf\endcsname
   \fam\sffam\csname sfIXf\endcsname}%
}%
\newfam\mibfam
\def\mib{%
   \getf@nt{mib}{IXf}{cmmib}{10 at9pt}%
   \getf@nt{mib}{VIf}{cmmib}{10 at6pt}%
   \getf@nt{mib}{Vf}{cmmib}{10 at5pt}%
   \textfont\mibfam=\csname mibIXf\endcsname
   \scriptfont\mibfam=\csname mibVIf\endcsname
   \scriptscriptfont\mibfam=\csname mibVf\endcsname
   \fam\mibfam\csname mibIXf\endcsname}%
\makesize{Xf}{10}{VIf}{6}{Vf}{5}{10.00dd}
\Xfsize
\it\bf\tt\rm

\def\tentt{\ttXf}

\normalsize
\it\bf\tt\sf\mib\rm
\def\boldmath{\textfont1=\mibIXf \scriptfont1=\mibVIf
\scriptscriptfont1=\mibVf}
\newdimen\fullhsize
\newcount\verybad \verybad=1010
\let\lr=L%
\fullhsize=40cc
\hsize=19.5cc
\def\fullline{\hbox to\fullhsize}
\def\makefootline{\baselineskip=10dd \fullline{\the\footline}}
\def\makeheadline{\vbox to 0pt{\vskip-22.5pt
            \fullline{\vbox to 8.5pt{}\the\headline}\vss}\nointerlineskip}
\hfuzz=2pt
\vfuzz=2pt
\tolerance=1000
\abovedisplayskip=3 mm plus6pt minus 4pt
\belowdisplayskip=3 mm plus6pt minus 4pt
\abovedisplayshortskip=0mm plus6pt
\belowdisplayshortskip=2 mm plus4pt minus 4pt
\parindent=1.5em
\newdimen\stdparindent\stdparindent\parindent
\frenchspacing
\nopagenumbers
\predisplaypenalty=600        
\displaywidowpenalty=2000     
\def\widowsandclubs#1{\global\verybad=#1
   \global\widowpenalty=\the\verybad1      
   \global\clubpenalty=\the\verybad2  }    
\widowsandclubs{1010}
\def\paglay{\headline={{\normalsize\hsize=.75\fullhsize\ifnum\pageno=1
\vbox{\hrule\line{\vrule\kern3pt\vbox{\kern3pt
\hbox{\bf A\&A manuscript no.}
\hbox{(will be inserted by hand later)}
\kern3pt\hrule\kern3pt
\hbox{\bf Your thesaurus codes are:}
\hbox{\rightskip=0pt plus3em\advance\hsize by-7pt
\vbox{\bf\noindent\ignorespaces\the\THESAURUS}}
\kern3pt}\hfil\kern3pt\vrule}\hrule}
\rlap{\quad\AALogo}\hfil
\else\normalsize\ifodd\pageno\hfil\folio\else\folio\hfil\fi\fi}}}
\makesize{VIIIf}{8}{VIf}{6}{Vf}{5}{9.00dd}
      \getf@nt{sf}{VIIIf}{cmss}{8}%
      \getf@nt{sf}{VIf}{cmss}{10 at 6pt}%
      \getf@nt{sf}{Vf}{cmss}{10 at 5pt}%
      \getf@nt{mib}{VIIIf}{cmmib}{10 at 8pt}%
      \getf@nt{mib}{VIf}{cmmib}{10 at 6pt}%
      \getf@nt{mib}{Vf}{cmmib}{10 at 5pt}%
\VIIIfsize\it\bf\tt\rm
\normalsize
\def\petit{\VIIIfsize
   \def\sf{%
      \getf@nt{sf}{VIIIf}{cmss}{8}%
      \getf@nt{sf}{VIf}{cmss}{10 at 6pt}%
      \getf@nt{sf}{Vf}{cmss}{10 at 5pt}%
      \textfont\sffam=\csname sfVIIIf\endcsname
      \scriptfont\sffam=\csname sfVIf\endcsname
      \scriptscriptfont\sffam=\csname sfVf\endcsname
      \fam\sffam\csname sfVIIIf\endcsname
}%
\def\mib{%
   \getf@nt{mib}{VIIIf}{cmmib}{10 at 8pt}%
   \getf@nt{mib}{VIf}{cmmib}{10 at 6pt}%
   \getf@nt{mib}{Vf}{cmmib}{10 at 5pt}%
   \textfont\mibfam=\csname mibVIIIf\endcsname
   \scriptfont\mibfam=\csname mibVIf\endcsname
   \scriptscriptfont\mibfam=\csname mibVf\endcsname
   \fam\mibfam\csname mibIXf\endcsname}%
\def\boldmath{\textfont1=\mibVIIIf\scriptfont1=\mibVIf
\scriptscriptfont1=\mibVf}%
\let\bfIXf=\bfVIIIf
 \if Y\REFEREE \normalbaselineskip=2\normalbaselineskip
 \normallineskip=2\normallineskip\fi
 \setbox\strutbox=\hbox{\vrule height7pt depth2pt width0pt}%
 \normalbaselines\rm}%
\def\begpet{\vskip\petitsurround
\bgroup\petit}
\def\endpet{\vskip\petitsurround
\egroup}

 \let  \tatss           = \bfXf
 \let  \tasss           = \syXf
 \let  \tbts            = \bfXf
 \let  \tbtss           = \bfVIIIf
 \let  \tbms            = \tamss
 \let  \tbmss           = \mibVIIIf
 \let  \tbss            = \syXf
 \let  \tbsss           = \syVIIIf
\def\newline{\hfill\break}
\def\rahmen#1{\vbox{\hrule\line{\vrule\vbox to#1true
cm{\vfil}\hfil\vrule}\vfil\hrule}}
\let\ts=\thinspace
\def\,{\relax\ifmmode\mskip\thinmuskip\else\thinspace\fi}
\def\unvskip{%
   \ifvmode
      \ifdim\lastskip=0pt
      \else
         \vskip-\lastskip
      \fi
   \fi}
\newtoks\eq\newtoks\eqn
\newdimen\mathhsize
\def\calcmathhsize{\mathhsize=\hsize
\advance\mathhsize by-\mathindent}
\calcmathhsize
\def\eqalign#1{\null\vcenter{\openup\jot\m@th
  \ialign{\strut\hfil$\displaystyle{##}$&$\displaystyle{{}##}$\hfil
      \crcr#1\crcr}}}
\def\displaylines#1{{}$\displ@y
\hbox{\vbox{\halign{$\@lign\hfil\displaystyle##\hfil$\crcr
    #1\crcr}}}${}}
\def\eqalignno#1{{}$\displ@y
  \hbox{\vbox{\halign
to\mathhsize{\hfil$\@lign\displaystyle{##}$\tabskip\z@skip
    &$\@lign\displaystyle{{}##}$\hfil\tabskip\centering
    &\llap{$\@lign##$}\tabskip\z@skip\crcr
    #1\crcr}}}${}}
\def\leqalignno#1{{}$\displ@y
\hbox{\vbox{\halign
to\mathhsize{\qquad\hfil$\@lign\displaystyle{##}$\tabskip\z@skip
    &$\@lign\displaystyle{{}##}$\hfil\tabskip\centering
    &\kern-\mathhsize\rlap{$\@lign##$}\tabskip\hsize\crcr
    #1\crcr}}}${}}
\def\generaldisplay{%
\ifeqno
       \ifleqno\leftline{$\displaystyle\the\eqn\quad\the\eq$}%
       \else\noindent\kern\mathindent\hbox to\mathhsize{$\displaystyle
             \the\eq\hfill\the\eqn$}%
       \fi
\else
       \kern\mathindent
       \hbox to\mathhsize{$\displaystyle\the\eq$\hss}%
\fi
\global\eq={}\global\eqn={}}%
\newif\ifeqno\newif\ifleqno
\everydisplay{\displaysetup}
\def\displaysetup#1$${\displaytest#1\eqno\eqno\displaytest}
\def\displaytest#1\eqno#2\eqno#3\displaytest{%
\if!#3!\ldisplaytest#1\leqno\leqno\ldisplaytest
\else\eqnotrue\leqnofalse\eqn={#2}\eq={#1}\fi
\generaldisplay$$}
\def\ldisplaytest#1\leqno#2\leqno#3\ldisplaytest{\eq={#1}%
\if!#3!\eqnofalse\else\eqnotrue\leqnotrue\eqn={#2}\fi}
\newcount\eqnum\eqnum=0
\def\autnum{\global\advance\eqnum by 1\relax{\rm(\the\eqnum)}}
\newdimen\lindent
\lindent=\stdparindent

\def\litemitem{\par\noindent\hbox to\lindent{\hfil}%
               \hangindent=2\lindent\ltextindent}
\def\ltextindent#1{\hbox to\lindent{#1\hss}\ignorespaces}
\def\set@item@mark#1{\llap{#1\enspace}\ignorespaces}
\ifx\undefined\mathhsize
   \def\item{\par\noindent
   \hangindent\itemindent\hangafter=0
   \set@item@mark}
   \def\itemitem{\par\noindent\advance\mathhsize by-\itemitemindent
   \hangindent\itemitemindent\hangafter=0
   \set@item@mark}
\else
   \def\item{\par\noindent\advance\mathhsize by-\itemindent
   \hangindent\itemindent\hangafter=0
   \everypar={\global\mathhsize=\hsize
   \global\advance\mathhsize by-\mathindent
   \global\everypar={}}\set@item@mark}
   \def\itemitem{\par\noindent\advance\mathhsize by-\itemitemindent
   \hangindent\itemitemindent\hangafter=0
   \everypar={\global\mathhsize=\hsize
   \global\advance\mathhsize by-\mathindent
   \global\everypar={}}\set@item@mark}
\fi
\newcount\the@end \global\the@end=0
\newbox\springer@macro \setbox\springer@macro=\vbox{}
\def\typeset{\setbox\springer@macro=\vbox{\begpet\noindent
   This article was processed by the author using
   Sprin\-ger-Ver\-lag \TeX{} A\&A macro package 1991.\par
   \egroup}\global\the@end=1}
\outer\def\bye{\bigskip\typeset
\sterne=1\ifx\speciali\undefined
\else
  \loop\smallskip\noindent special character No\number\sterne:
    \csname special\romannumeral\sterne\endcsname
    \advance\sterne by 1\relax
    \ifnum\sterne<11\relax
  \repeat
\fi
\if R\lr\null\fi\vfill\supereject\end}
\def\AALogo{\setbox254=\hbox{ ASTROPHYSICS }%
\vbox{\baselineskip=10dd\hrule\hbox{\vrule\vbox{\kern3pt
\hbox to\wd254{\hfil ASTRONOMY\hfil}
\hbox to\wd254{\hfil AND\hfil}\copy254
\hbox to\wd254{\hfil\number\day.\number\month.\number\year\hfil}
\kern3pt}\vrule}\hrule}}
\def\figure#1#2{\medskip\noindent{\petit{\bf Fig.\ts#1.\
}\ignorespaces#2\par}}
\expandafter \newcount \csname c@Tl\endcsname
    \csname c@Tl\endcsname=0
\expandafter \newcount \csname c@Tm\endcsname
    \csname c@Tm\endcsname=0
\expandafter \newcount \csname c@Tn\endcsname
    \csname c@Tn\endcsname=0
\expandafter \newcount \csname c@To\endcsname
    \csname c@To\endcsname=0
\expandafter \newcount \csname c@Tp\endcsname
    \csname c@Tp\endcsname=0
\expandafter \newcount \csname c@fn\endcsname
    \csname c@fn\endcsname=0
\def \stepc#1    {\global
    \expandafter
    \advance
    \csname c@#1\endcsname by 1}
\def \resetcount#1    {\global
    \csname c@#1\endcsname=0}
\def\@nameuse#1{\csname #1\endcsname}
\def\arabic#1{\@arabic{\@nameuse{c@#1}}}
\def\@arabic#1{\ifnum #1>0 \number #1\fi}
 \def \aTa  { \goodbreak
     \bgroup
     \par
 \textfont0=\tatt \scriptfont0=\tats \scriptscriptfont0=\tatss
 \textfont1=\tamt \scriptfont1=\tams \scriptscriptfont1=\tamss
 \textfont2=\tast \scriptfont2=\tass \scriptscriptfont2=\tasss
     \baselineskip=17dd\lineskiplimit=0pt\lineskip=0pt
     \rightskip=0pt plus4cm
     \pretolerance=10000
     \noindent
     \tatt}
 \def \eTa{\vskip10pt\egroup
     \noindent
     \ignorespaces}
 \def \aTb{\goodbreak
     \bgroup
     \par
 \textfont0=\tbtt \scriptfont0=\tbts \scriptscriptfont0=\tbtss
 \textfont1=\tbmt \scriptfont1=\tbms \scriptscriptfont1=\tbmss
 \textfont2=\tbst \scriptfont2=\tbss \scriptscriptfont2=\tbsss
     \baselineskip=13dd\lineskip=0pt\lineskiplimit=0pt
     \rightskip=0pt plus4cm
     \pretolerance=10000
     \noindent
     \tbtt}
 \def \eTb{\vskip10pt
     \egroup
     \noindent
     \ignorespaces}
\newcount\section@penalty  \section@penalty=0
\newcount\subsection@penalty  \subsection@penalty=0
\newcount\subsubsection@penalty  \subsubsection@penalty=0
\def\titlea#1{\par\stepc{Tl}
    \resetcount{Tm}
    \bgroup
       \normalsize
       \bf \rightskip 0pt plus4em
       \pretolerance=20000
       \boldmath
       \setbox0=\vbox{\vskip\tabefore
          \noindent
          \arabic{Tl}.\
          \ignorespaces#1
          \vskip\taafter}
       \dimen0=\ht0\advance\dimen0 by\dp0
       \advance\dimen0 by 2\baselineskip
       \advance\dimen0 by\pagetotal
       \ifdim\dimen0>\pagegoal
          \ifdim\pagetotal>\pagegoal
          \else\eject\fi\fi
       \vskip\tabefore
       \penalty\section@penalty \global\section@penalty=-200
       \global\subsection@penalty=10007
       \noindent
       \arabic{Tl}.\
       \ignorespaces#1
       \vskip\taafter
    \egroup
    \nobreak
    \parindent=0pt
    \let\lasttitle=A%
\everypar={\parindent=\stdparindent
    \penalty\z@\let\lasttitle=N\everypar={}}%
    \ignorespaces}
\def\titleb#1{\par\stepc{Tm}
    \resetcount{Tn}
    \if N\lasttitle\else\vskip\tbbeforeback\fi
    \bgroup
       \normalsize
       \raggedright
       \pretolerance=10000
       \it
       \setbox0=\vbox{\vskip\tbbefore
          \normalsize
          \raggedright
          \pretolerance=10000
          \noindent \it \arabic{Tl}.\arabic{Tm}.\ \ignorespaces#1
          \vskip\tbafter}
       \dimen0=\ht0\advance\dimen0 by\dp0\advance\dimen0 by 2\baselineskip
       \advance\dimen0 by\pagetotal
       \ifdim\dimen0>\pagegoal
          \ifdim\pagetotal>\pagegoal
          \else \if N\lasttitle\eject\fi \fi\fi
       \vskip\tbbefore
       \if N\lasttitle \penalty\subsection@penalty \fi
       \global\subsection@penalty=-100
       \global\subsubsection@penalty=10007
       \noindent \arabic{Tl}.\arabic{Tm}.\ \ignorespaces#1
       \vskip\tbafter
    \egroup
    \nobreak
    \let\lasttitle=B%
    \parindent=0pt
    \everypar={\parindent=\stdparindent
       \penalty\z@\let\lasttitle=N\everypar={}}%
       \ignorespaces}
\def\titlec#1{\par\stepc{Tn}
    \resetcount{To}
    \if N\lasttitle\else\vskip\tcbeforeback\fi
    \bgroup
       \normalsize
       \raggedright
       \pretolerance=10000
       \setbox0=\vbox{\vskip\tcbefore
          \noindent
          \arabic{Tl}.\arabic{Tm}.\arabic{Tn}.\
          \ignorespaces#1\vskip\tcafter}
       \dimen0=\ht0\advance\dimen0 by\dp0\advance\dimen0 by 2\baselineskip
       \advance\dimen0 by\pagetotal
       \ifdim\dimen0>\pagegoal
           \ifdim\pagetotal>\pagegoal
           \else \if N\lasttitle\eject\fi \fi\fi
       \vskip\tcbefore
       \if N\lasttitle \penalty\subsubsection@penalty \fi
       \global\subsubsection@penalty=-50
       \noindent
       \arabic{Tl}.\arabic{Tm}.\arabic{Tn}.\
       \ignorespaces#1\vskip\tcafter
    \egroup
    \nobreak
    \let\lasttitle=C%
    \parindent=0pt
    \everypar={\parindent=\stdparindent
       \penalty\z@\let\lasttitle=N\everypar={}}%
       \ignorespaces}
\def\titled#1{\par\stepc{To}
    \resetcount{Tp}
    \if N\lasttitle\else\vskip\tdbeforeback\fi
    \vskip\tdbefore
    \bgroup
       \normalsize
       \if N\lasttitle \penalty-50 \fi
       \it \noindent \ignorespaces#1\unskip\
    \egroup\ignorespaces}
\def\begref#1{\par
   \unvskip
   \goodbreak\vskip\tabefore
   {\noindent\bf\ignorespaces#1%
   \par\vskip\taafter}\nobreak\let\INS=N}
\def\ref{\if N\INS\let\INS=Y\else\goodbreak\fi
   \hangindent\stdparindent\hangafter=1\noindent\ignorespaces}
\def\endref{\goodbreak}
\def\acknow#1{\par
   \unvskip
   \vskip\tcbefore
   \noindent{\it Acknowledgements\/}. %
   \ignorespaces#1\par
   \vskip\tcafter}
\def\appendix#1{\vskip\tabefore
    \vbox{\noindent{\bf Appendix #1}\vskip\taafter}%
    \global\eqnum=0\relax
    \nobreak\noindent\ignorespaces}
\let\REFEREE=N
\newbox\refereebox
\setbox\refereebox=\vbox
to0pt{\vskip0.5cm\fullline{\hrulefill\tentt\lower0.5ex
\hbox{\kern5pt referee's copy\kern5pt}\hrulefill}\vss}%
\def\refereelayout{\let\REFEREE=M\footline={\copy\refereebox}
    \message{|A referee's copy will be produced}\par
    \if N\lr\else\if R\lr \onecolumn \fi \let\lr=N \topskip=10pt\fi}

\def\utw{\smash{\rlap{\lower5pt\hbox{$\sim$}}}}
\def\udtw{\smash{\rlap{\lower6pt\hbox{$\approx$}}}}


\def\bbbc{{\mathchoice {\setbox0=\hbox{$\displaystyle\rm C$}\hbox{\hbox
to0pt{\kern0.4\wd0\vrule height0.9\ht0\hss}\box0}}
{\setbox0=\hbox{$\textstyle\rm C$}\hbox{\hbox
to0pt{\kern0.4\wd0\vrule height0.9\ht0\hss}\box0}}
{\setbox0=\hbox{$\scriptstyle\rm C$}\hbox{\hbox
to0pt{\kern0.4\wd0\vrule height0.9\ht0\hss}\box0}}
{\setbox0=\hbox{$\scriptscriptstyle\rm C$}\hbox{\hbox
to0pt{\kern0.4\wd0\vrule height0.9\ht0\hss}\box0}}}}
\def\bbbq{{\mathchoice {\setbox0=\hbox{$\displaystyle\rm Q$}\hbox{\raise
0.15\ht0\hbox to0pt{\kern0.4\wd0\vrule height0.8\ht0\hss}\box0}}
{\setbox0=\hbox{$\textstyle\rm Q$}\hbox{\raise
0.15\ht0\hbox to0pt{\kern0.4\wd0\vrule height0.8\ht0\hss}\box0}}
{\setbox0=\hbox{$\scriptstyle\rm Q$}\hbox{\raise
0.15\ht0\hbox to0pt{\kern0.4\wd0\vrule height0.7\ht0\hss}\box0}}
{\setbox0=\hbox{$\scriptscriptstyle\rm Q$}\hbox{\raise
0.15\ht0\hbox to0pt{\kern0.4\wd0\vrule height0.7\ht0\hss}\box0}}}}
\def\bbbt{{\mathchoice {\setbox0=\hbox{$\displaystyle\rm
T$}\hbox{\hbox to0pt{\kern0.3\wd0\vrule height0.9\ht0\hss}\box0}}
{\setbox0=\hbox{$\textstyle\rm T$}\hbox{\hbox
to0pt{\kern0.3\wd0\vrule height0.9\ht0\hss}\box0}}
{\setbox0=\hbox{$\scriptstyle\rm T$}\hbox{\hbox
to0pt{\kern0.3\wd0\vrule height0.9\ht0\hss}\box0}}
{\setbox0=\hbox{$\scriptscriptstyle\rm T$}\hbox{\hbox
to0pt{\kern0.3\wd0\vrule height0.9\ht0\hss}\box0}}}}
\def\bbbs{{\mathchoice
{\setbox0=\hbox{$\displaystyle     \rm S$}\hbox{\raise0.5\ht0\hbox
to0pt{\kern0.35\wd0\vrule height0.45\ht0\hss}\hbox
to0pt{\kern0.55\wd0\vrule height0.5\ht0\hss}\box0}}
{\setbox0=\hbox{$\textstyle        \rm S$}\hbox{\raise0.5\ht0\hbox
to0pt{\kern0.35\wd0\vrule height0.45\ht0\hss}\hbox
to0pt{\kern0.55\wd0\vrule height0.5\ht0\hss}\box0}}
{\setbox0=\hbox{$\scriptstyle      \rm S$}\hbox{\raise0.5\ht0\hbox
to0pt{\kern0.35\wd0\vrule height0.45\ht0\hss}\raise0.05\ht0\hbox
to0pt{\kern0.5\wd0\vrule height0.45\ht0\hss}\box0}}
{\setbox0=\hbox{$\scriptscriptstyle\rm S$}\hbox{\raise0.5\ht0\hbox
to0pt{\kern0.4\wd0\vrule height0.45\ht0\hss}\raise0.05\ht0\hbox
to0pt{\kern0.55\wd0\vrule height0.45\ht0\hss}\box0}}}}
\def\bbbz{{\mathchoice {\hbox{$\sf\textstyle Z\kern-0.4em Z$}}
{\hbox{$\sf\textstyle Z\kern-0.4em Z$}}
{\hbox{$\sf\scriptstyle Z\kern-0.3em Z$}}
{\hbox{$\sf\scriptscriptstyle Z\kern-0.2em Z$}}}}
\def\diameter{{\ifmmode\oslash\else$\oslash$\fi}}

\def\vec#1{{\boldmath
\textfont0=\bfIXf\scriptfont0=\bfVIf\scriptscriptfont0=\bfVf
\ifmmode
\mathchoice{\hbox{$\displaystyle#1$}}{\hbox{$\textstyle#1$}}
{\hbox{$\scriptstyle#1$}}{\hbox{$\scriptscriptstyle#1$}}\else
$#1$\fi}}
\def\tens#1{\ifmmode
\mathchoice{\hbox{$\displaystyle\sf#1$}}{\hbox{$\textstyle\sf#1$}}
{\hbox{$\scriptstyle\sf#1$}}{\hbox{$\scriptscriptstyle\sf#1$}}\else
$\sf#1$\fi}
\newcount\sterne \sterne=0
\newdimen\fullhead
{\catcode`@=11    
\def\newtoks{\alloc@5\toks\toksdef\@cclvi}
\outer\gdef\makenewtoks#1{\newtoks#1#1={ ????? }}}
\makenewtoks\DATE
\makenewtoks\MAINTITLE
\makenewtoks\SUBTITLE
\makenewtoks\AUTHOR
\makenewtoks\INSTITUTE
\makenewtoks\ABSTRACT
\makenewtoks\KEYWORDS
\makenewtoks\THESAURUS
\makenewtoks\OFFPRINTS
\newlinechar=`\| %
\let\INS=N%
{\catcode`\@=\active
\gdef@#1{\if N\INS $^{#1}$\else\if
E\INS\hangindent0.5\stdparindent\hangafter=1%
\noindent\hbox to0.5\stdparindent{$^{#1}$\hfil}\let\INS=Y\ignorespaces
\else\par\hangindent0.5\stdparindent\hangafter=1
\noindent\hbox to0.5\stdparindent{$^{#1}$\hfil}\ignorespaces\fi\fi}%
}%
\def\mehrsterne{\global\advance\sterne by1\relax}%
\def\footnoterule{\kern-3pt\hrule width 2true cm\kern2.6pt}
\def\makeOFFPRINTS#1{\bgroup\normalsize
       \hsize=19.5cc
       \baselineskip=10dd\lineskiplimit=0pt\lineskip=0pt
       \def\textindent##1{\noindent{\it Send offprint
          requests to\/}: }\relax
       \vfootnote{nix}{\ignorespaces#1}\egroup}
\def\makesterne{\count254=0\loop\ifnum\count254<\sterne
\advance\count254 by1\star\repeat}
\def\FOOTNOTE#1{\bgroup
       \ifhmode\unskip\fi
       \mehrsterne$^{\makesterne}$\relax
       \normalsize
       \hsize=19.5cc
       \baselineskip=10dd\lineskiplimit=0pt\lineskip=0pt
       \def\textindent##1{\noindent\hbox
       to\stdparindent{##1\hss}}\relax
       \vfootnote{$^{\makesterne}$}{\ignorespaces#1}\egroup}
\def\fonote#1{\ifhmode\unskip\fi
       \mehrsterne$^{\the\sterne}$\bgroup
       \normalsize
       \hsize=19.5cc
       \def\textindent##1{\noindent\hbox
       to\stdparindent{##1\hss}}\relax
       \vfootnote{$^{\the\sterne}$}{\ignorespaces#1}\egroup}
\def\missmsg#1{\message{|Missing #1 }}
\def\tstmiss#1#2#3#4#5{%
\edef\test{\the #1}%
\ifx\test\missing%
  #2\relax
  #3
\else
  \ifx\test\missingi%
    #2\relax
    #3
  \else #4
  \fi
\fi
#5
}%
\def\maketitle{\paglay%
\def\missing{ ????? }%
\def\missingi{ }%
{\parskip=0pt\relax
\setbox0=\vbox{\hsize=\fullhsize\null\vskip2truecm
\tstmiss%
  {\MAINTITLE}%
  {}%
  {\global\MAINTITLE={MAINTITLE should be given}}%
  {}%
  {
   \aTa\ignorespaces\the\MAINTITLE\eTa}%
\tstmiss%
  {\SUBTITLE}%
  {}%
  {}%
  {
   \aTb\ignorespaces\the\SUBTITLE\eTb}%
  {}%
\tstmiss%
  {\AUTHOR}%
  {}%
  {\AUTHOR={Name(s) and initial(s) of author(s) should be given}}
  {}%
  {
\noindent{\bf\ignorespaces\the\AUTHOR\vskip4pt}}%
\tstmiss%
  {\INSTITUTE}%
  {}%
  {\INSTITUTE={Address(es) of author(s) should be given.}}%
  {}%
  {
   \let\INS=E
\noindent\ignorespaces\the\INSTITUTE\vskip10pt}%
\tstmiss%
  {\DATE}%
  {}%
  {\DATE={$[$the date of receipt and acceptance should be inserted
later$]$}}%
  {}%
  {
{\noindent\ignorespaces\the\DATE\vskip21pt}\bf A}%
}%
\global\fullhead=\ht0\global\advance\fullhead by\dp0
\global\advance\fullhead by10pt\global\sterne=0
{\hsize=19.5cc\null\vskip2truecm
\tstmiss%
  {\OFFPRINTS}%
  {}%
  {}%
  {\makeOFFPRINTS{\the\OFFPRINTS}}%
  {}%
\hsize=\fullhsize
\tstmiss%
  {\MAINTITLE}%
  {\missmsg{MAINTITLE}}%
  {\global\MAINTITLE={MAINTITLE should be given}}%
  {}%
  {
   \aTa\ignorespaces\the\MAINTITLE\eTa}%
\tstmiss%
  {\SUBTITLE}%
  {}%
  {}%
  {
   \aTb\ignorespaces\the\SUBTITLE\eTb}%
  {}%
\tstmiss%
  {\AUTHOR}%
  {\missmsg{name(s) and initial(s) of author(s)}}%
  {\AUTHOR={Name(s) and initial(s) of author(s) should be given}}
  {}%
  {
\noindent{\bf\ignorespaces\the\AUTHOR\vskip4pt}}%
\tstmiss%
  {\INSTITUTE}%
  {\missmsg{address(es) of author(s)}}%
  {\INSTITUTE={Address(es) of author(s) should be given.}}%
  {}%
  {
   \let\INS=E
\noindent\ignorespaces\the\INSTITUTE\vskip10pt}%
\catcode`\@=12
\tstmiss%
  {\DATE}%
  {\message{|The date of receipt and acceptance should be inserted
later.}}%
  {\DATE={$[$the date of receipt and acceptance should be inserted
later$]$}}%
  {}%
  {
{\noindent\ignorespaces\the\DATE\vskip21pt}}%
}%
\tstmiss%
  {\THESAURUS}%
  {\message{|Thesaurus codes are not given.}}%
  {\global\THESAURUS={missing; you have not inserted them}}%
  {}%
  {}%
\if M\REFEREE\let\REFEREE=Y
\normalbaselineskip=2\normalbaselineskip
\normallineskip=2\normallineskip\normalbaselines\fi
\tstmiss%
  {\ABSTRACT}%
  {\missmsg{ABSTRACT}}%
  {\ABSTRACT={Not yet given.}}%
  {}%
  {\noindent{\bf Abstract. }\ignorespaces\the\ABSTRACT\vskip0.5true cm}%
\def\strich{\par
\vbox to0pt{\hrule width\hsize\vss}\vskip-1.2\baselineskip
\vskip0pt plus3\baselineskip\relax}%
\tstmiss%
  {\KEYWORDS}%
  {\missmsg{KEYWORDS}}%
  {\KEYWORDS={Not yet given.}}%
  {}%
  {\noindent{\bf Key words: }\ignorespaces\the\KEYWORDS
  \strich}%
\global\sterne=0
}}
\newdimen\@txtwd  \@txtwd=\hsize
\newdimen\@txtht  \@txtht=\vsize
\newdimen\@colht  \@colht=\vsize
\newdimen\@colwd  \@colwd=-1pt
\newdimen\@colsavwd
\newcount\in@t \in@t=0
\def\initlr{\if N\lr \ifdim\@colwd<0pt \global\@colwd=\hsize \fi
   \else\global\let\lr=L\ifdim\@colwd<0pt \global\@colwd=\hsize
      \global\divide\@colwd\tw@ \global\advance\@colwd by -10pt
   \fi\fi\global\advance\in@t by 1}
\def\setuplr#1#2#3{\let\lr=O \ifx#1\lr\global\let\lr=N
      \else\global\let\lr=L\fi
   \@txtht=\vsize \@colht=\vsize \@txtwd=#2 \@colwd=#3
   \if N\lr \else\multiply\@colwd\tw@ \fi
   \ifdim\@colwd>\@txtwd\if N\lr
        \errmessage{The text width is less than the column width}%
      \else
        \errmessage{The text width is less the two times the column width}%
      \fi \global\@colwd=\@txtwd
      \if N\lr\divide\@colwd by 2\fi
   \else \global\@colwd=#3 \fi \initlr \@colsavwd=#3
   \global\@insmx=\@txtht
   \global\hsize=\@colwd}
\def\twocolumns{\@fillpage\eject\global\let\lr=L \@makecolht
   \global\@colwd=\@colsavwd \global\hsize=\@colwd}
\def\onecolumn{\@fillpage\eject\global\let\lr=N \@makecolht
   \global\@colwd=\@txtwd \global\hsize=\@colwd}
\def\newpage{\@fillpage\eject}
\def\@fillpage{\vfill\supereject\if R\lr \null\vfill\eject\fi}

\newbox\@leftcolumn
\newbox\@rightcolumn
\newbox\@outputbox
\newbox\@tempboxa
\newbox\@keepboxa
\newbox\@keepboxb
\newbox\@bothcolumns
\newbox\@savetopins
\newbox\@savetopright
\newcount\verybad \verybad=1010
\def\@makecolumn{\ifnum \in@t<1\initlr\fi
   \ifnum\outputpenalty=\the\verybad1  
      \if L\lr\else\advance\pageno by1\fi
      \message{Warning: There is a 'widow' line
      at the top of page \the\pageno\if R\lr (left)\fi.
      This is unacceptable.} \if L\lr\else\advance\pageno by-1\fi \fi
   \ifnum\outputpenalty=\the\verybad2
      \message{Warning: There is a 'club' line
      at the bottom of page \the\pageno\if L\lr(left)\fi.
      This is unacceptable.} \fi
   \if L\lr \ifvoid\@savetopins\else\@colht=\@txtht\fi \fi
   \if R\lr \ifvoid\@bothcolumns \ifvoid\@savetopright
       \else\@colht=\@txtht\fi\fi\fi
   \global\setbox\@outputbox
   \vbox to\@colht{\boxmaxdepth\maxdepth
   \if L\lr \ifvoid\@savetopins\else\unvbox\@savetopins\fi \fi
   \if R\lr \ifvoid\@bothcolumns \ifvoid\@savetopright\else
       \unvbox\@savetopright\fi\fi\fi
   \ifvoid\topins\else\ifnum\count\topins>0
         \ifdim\ht\topins>\@colht
            \message{|Error: Too many or too large single column
            box(es) on this page.}\fi
         \unvbox\topins
      \else
         \global\setbox\@savetopins=\vbox{\ifvoid\@savetopins\else
         \unvbox\@savetopins\penalty-500\fi \unvbox\topins} \fi\fi
   \dimen@=\dp\@cclv \unvbox\@cclv 
   \ifvoid\bottomins\else\unvbox\bottomins\fi
   \ifvoid\footins\else 
     \vskip\skip\footins
     \footnoterule
     \unvbox\footins\fi
   \ifr@ggedbottom \kern-\dimen@ \vfil \fi}%
}
\def\@outputpage{\@dooutput{\lr}}
\def\@colbox#1{\hbox to\@colwd{\box#1\hss}}
\def\@dooutput#1{\global\topskip=10pt
  \ifdim\ht\@bothcolumns>\@txtht
    \if #1N
       \unvbox\@outputbox
    \else
       \unvbox\@leftcolumn\unvbox\@outputbox
    \fi
    \global\setbox\@tempboxa\vbox{\hsize=\@txtwd\makeheadline
       \vsplit\@bothcolumns to\@txtht
       \makefootline\hsize=\@colwd}%
    \message{|Error: Too many double column boxes on this page.}%
    \shipout\box\@tempboxa\advancepageno
    \unvbox255 \penalty\outputpenalty
  \else
    \global\setbox\@tempboxa\vbox{\hsize=\@txtwd\makeheadline
       \ifvoid\@bothcolumns\else\unvbox\@bothcolumns\fi
       \hsize=\@colwd
       \if #1N
          \hbox to\@txtwd{\@colbox{\@outputbox}\hfil}%
       \else
          \hbox to\@txtwd{\@colbox{\@leftcolumn}\hfil\@colbox{\@outputbox}}%
       \fi
       \hsize=\@txtwd\makefootline\hsize=\@colwd}%
    \shipout\box\@tempboxa\advancepageno
  \fi
  \ifnum \special@pages>0 \s@count=100 \page@command
      \xdef\page@command{}\global\special@pages=0 \fi
  }
\def\balance@right@left{\dimen@=\ht\@leftcolumn
    \advance\dimen@ by\ht\@outputbox
    \advance\dimen@ by\ht\springer@macro
    \dimen2=\z@ \global\the@end=0
    \ifdim\dimen@>70pt\setbox\z@=\vbox{\unvbox\@leftcolumn
          \unvbox\@outputbox}%
       \loop
          \dimen@=\ht\z@
          \advance\dimen@ by0.5\topskip
          \advance\dimen@ by\baselineskip
          \advance\dimen@ by\ht\springer@macro
          \advance\dimen@ by\dimen2
          \divide\dimen@ by2
          \splittopskip=\topskip
          {\vbadness=10000
             \global\setbox3=\copy\z@
             \global\setbox1=\vsplit3 to \dimen@}%
          \dimen1=\ht3 \advance\dimen1 by\ht\springer@macro
       \ifdim\dimen1>\ht1 \advance\dimen2 by\baselineskip\repeat
       \dimen@=\ht1
       \global\setbox\@leftcolumn
          \hbox to\@colwd{\vbox to\@colht{\vbox to\dimen@{\unvbox1}\vfil}}%
       \global\setbox\@outputbox
          \hbox to\@colwd{\vbox to\@colht{\vbox to\dimen@{\unvbox3
             \vfill\box\springer@macro}\vfil}}%
    \else
       \setbox\@leftcolumn=\vbox{unvbox\@leftcolumn\bigskip
          \box\springer@macro}%
    \fi}
\newinsert\bothins
\newbox\rightins
\skip\bothins=\z@skip
\count\bothins=1000
\dimen\bothins=\@txtht \advance\dimen\bothins by -\bigskipamount
\def\bothtopinsert{\par\begingroup\setbox\z@\vbox\bgroup
    \hsize=\@txtwd\parskip=0pt\par\noindent\bgroup}
\def\endbothinsert{\egroup\egroup
  \if R\lr
    \right@nsert
  \else    
    \dimen@=\ht\z@ \advance\dimen@ by\dp\z@ \advance\dimen@ by\pagetotal
    \advance\dimen@ by \bigskipamount \advance\dimen@ by \topskip
    \advance\dimen@ by\ht\topins \advance\dimen@ by\dp\topins
    \advance\dimen@ by\ht\bottomins \advance\dimen@ by\dp\bottomins
    \advance\dimen@ by\ht\@savetopins \advance\dimen@ by\dp\@savetopins
    \ifdim\dimen@>\@colht\right@nsert\else\left@nsert\fi
  \fi  \endgroup}
\def\right@nsert{\global\setbox\rightins\vbox{\ifvoid\rightins
    \else\unvbox\rightins\fi\penalty100
    \splittopskip=\topskip
    \splitmaxdepth\maxdimen \floatingpenalty200
    \dimen@\ht\z@ \advance\dimen@\dp\z@
    \box\z@\nobreak\bigskip}}
\def\left@nsert{\insert\bothins{\penalty100
    \splittopskip=\topskip
    \splitmaxdepth\maxdimen \floatingpenalty200
    \box\z@\nobreak\bigskip}
    \@makecolht}
\newdimen\@insht    \@insht=\z@
\newdimen\@insmx    \@insmx=\vsize
\def\@makecolht{\global\@colht=\@txtht \@compinsht
    \global\advance\@colht by -\@insht \global\vsize=\@colht
    \global\dimen\topins=\@colht}
\def\@compinsht{\if R\lr
       \dimen@=\ht\@bothcolumns \advance\dimen@ by\dp\@bothcolumns
       \ifvoid\@bothcolumns \advance\dimen@ by\ht\@savetopright
          \advance\dimen@ by\dp\@savetopright \fi
    \else
       \dimen@=\ht\bothins \advance\dimen@ by\dp\bothins
       \advance\dimen@ by\ht\@savetopins \advance\dimen@ by\dp\@savetopins
    \fi
    \ifdim\dimen@>\@insmx
       \global\@insht=\dimen@
    \else\global\@insht=\dimen@
    \fi}
\newinsert\bottomins
\skip\bottomins=\z@skip
\count\bottomins=1000
\xdef\page@command{}
\newcount\s@count
\newcount\special@pages \special@pages=0
\def\specialpage#1{\global\advance\special@pages by1
    \global\s@count=\special@pages
    \global\advance\s@count by 100
    \global\setbox\s@count
    \vbox to\@txtht{\hsize=\@txtwd\parskip=0pt
    \par\noindent\noexpand#1\vfil}%
    \def\protect{\noexpand\protect\noexpand}%
    \xdef\page@command{\page@command
         \protect\global\advance\s@count by1
         \protect\begingroup
         \protect\setbox\z@\vbox{\protect\makeheadline
                                    \protect\box\s@count
            \protect\makefootline}%
         \protect{\shipout\box\z@}%
         \protect\endgroup\protect\advancepageno}%
    \let\protect=\relax
   }
\def\@startins{\vskip \topskip\hrule height\z@
   \nobreak\vskip -\topskip\vskip3.7pt}
\let\retry=N
\output={\@makecolht \global\topskip=10pt \let\retry=N%
   \ifnum\count\topins>0 \ifdim\ht\topins>\@colht
       \global\count\topins=0 \global\let\retry=Y%
       \unvbox\@cclv \penalty\outputpenalty \fi\fi
   \if N\retry
    \if N\lr     
       \@makecolumn
       \ifnum\the@end>0
          \setbox\z@=\vbox{\unvcopy\@outputbox}%
          \dimen@=\ht\z@ \advance\dimen@ by\ht\springer@macro
          \ifdim\dimen@<\@colht
             \setbox\@outputbox=\vbox to\@colht{\box\z@
             \unskip\vskip12pt plus0pt minus12pt
             \box\springer@macro\vfil}%
          \else \box\springer@macro \fi
          \global\the@end=0
       \fi
       \ifvoid\bothins\else\global\setbox\@bothcolumns\box\bothins\fi
       \@outputpage
       \ifvoid\rightins\else
       \ifvoid\@savetopins\insert\bothins{\unvbox\rightins}\fi
       \fi
    \else
       \if L\lr    
          \@makecolumn
          \global\setbox\@leftcolumn\box\@outputbox \global\let\lr=R%
          \ifnum\pageno=1
             \message{|[left\the\pageno]}%
          \else
             \message{[left\the\pageno]}\fi
          \ifvoid\bothins\else\global\setbox\@bothcolumns\box\bothins\fi
          \global\dimen\bothins=\z@
          \global\count\bothins=0
          \ifnum\pageno=1
             \global\topskip=\fullhead\fi
       \else    
          \@makecolumn
          \ifnum\the@end>0\ifnum\pageno>1\balance@right@left\fi\fi
          \@outputpage \global\let\lr=L%
          \global\dimen\bothins=\maxdimen
          \global\count\bothins=1000
          \ifvoid\rightins\else
             \ifvoid\@savetopins \insert\bothins{\unvbox\rightins}\fi
          \fi
       \fi
    \fi
    \global\let\last@insert=N \put@default
    \ifnum\outputpenalty>-\@MM\else\dosupereject\fi
    \ifvoid\@savetopins\else
      \ifdim\ht\@savetopins>\@txtht
        \global\setbox\@tempboxa=\box\@savetopins
        \global\setbox\@savetopins=\vsplit\@tempboxa to\@txtht
        \global\setbox\@savetopins=\vbox{\unvbox\@savetopins}%
        \global\setbox\@savetopright=\box\@tempboxa \fi
    \fi
    \@makecolht
    \global\count\topins=1000
   \fi
   }
\if N\lr
   \setuplr{O}{\fullhsize}{\hsize}
\else
   \setuplr{T}{\fullhsize}{\hsize}
\fi
\def\put@default{\global\let\insert@here=Y
   \global\let\insert@at@the@bottom=N}%
\def\puthere{\global\let\insert@here=Y%
    \global\let\insert@at@the@bottom=N}
\def\putattop{\global\let\insert@here=N%
    \global\let\insert@at@the@bottom=N}
\def\putatbottom{\global\let\insert@here=N%
    \global\let\insert@at@the@bottom=X}
\put@default
\let\last@insert=N
\def\end@skip{\smallskip}
\newdimen\min@top
\newdimen\min@here
\newdimen\min@bot
\min@top=10cm
\min@here=4cm
\min@bot=\topskip
\def\figfuzz{\vskip 0pt plus 6pt minus 3pt}  
\def\check@here@and@bottom#1{\relax
   \ifvoid\topins\else       \global\let\insert@here=N\fi
   \if B\last@insert         \global\let\insert@here=N\fi
   \if T\last@insert         \global\let\insert@here=N\fi
   \ifdim #1<\min@bot        \global\let\insert@here=N\fi
   \ifdim\pagetotal>\@colht  \global\let\insert@here=N\fi
   \ifdim\pagetotal<\min@here\global\let\insert@here=N\fi
   \if X\insert@at@the@bottom\global\let\insert@at@the@bottom=Y
     \else\if T\last@insert  \global\let\insert@at@the@bottom=N\fi
          \if H\last@insert  \global\let\insert@at@the@bottom=N\fi
          \ifvoid\topins\else\global\let\insert@at@the@bottom=N\fi\fi
   \ifdim #1<\min@bot        \global\let\insert@at@the@bottom=N\fi
   \ifdim\pagetotal>\@colht  \global\let\insert@at@the@bottom=N\fi
   \ifdim\pagetotal<\min@top \global\let\insert@at@the@bottom=N\fi
   \ifvoid\bottomins\else    \global\let\insert@at@the@bottom=Y\fi
   \if Y\insert@at@the@bottom\global\let\insert@here=N\fi }
\def\single@column@insert#1{\relax
   \setbox\@tempboxa=\vbox{#1}%
   \dimen@=\@colht \advance\dimen@ by -\pagetotal
   \advance\dimen@ by-\ht\@tempboxa \advance\dimen0 by-\dp\@tempboxa
   \advance\dimen@ by-\ht\topins \advance\dimen0 by-\dp\topins
   \check@here@and@bottom{\dimen@}%
   \if Y\insert@here
      \par  
      \midinsert\figfuzz\relax     
      \box\@tempboxa\end@skip\figfuzz\endinsert
      \global\let\last@insert=H
   \else \if Y\insert@at@the@bottom
      \begingroup\insert\bottomins\bgroup\if B\last@insert\end@skip\fi
      \floatingpenalty=20000\figfuzz\bigskip\box\@tempboxa\egroup\endgroup
      \global\let\last@insert=B
   \else
      \topinsert\box\@tempboxa\end@skip\figfuzz\endinsert
      \global\let\last@insert=T
   \fi\fi\put@default\ignorespaces}
\def\begfig#1cm#2\endfig{\single@column@insert{\@startins\rahmen{#1}#2}%
\ignorespaces}
\def\begfigwid#1cm#2\endfig{\relax
   \if N\lr  
      {\hsize=\fullhsize \begfig#1cm#2\endfig}%
   \else
      \setbox0=\vbox{\hsize=\fullhsize\bigskip#2\smallskip}%
      \dimen0=\ht0\advance\dimen0 by\dp0
      \advance\dimen0 by#1cm
      \advance\dimen0by7\normalbaselineskip\relax
      \ifdim\dimen0>\@txtht
         \message{|Figure plus legend too high, will try to put it on a
                  separate page. }%
         \begfigpage#1cm#2\endfig
      \else
         \bothtopinsert\line{\vbox{\hsize=\fullhsize
         \@startins\rahmen{#1}#2\smallskip}\hss}\figfuzz\endbothinsert
      \fi
   \fi}
\def\begfigside#1cm#2cm#3\endfig{\relax
   \if N\lr  
      {\hsize=\fullhsize \begfig#1cm#3\endfig}%
   \else
      \dimen0=#2true cm\relax
      \ifdim\dimen0<\hsize
         \message{|Your figure fits in a single column; why don't|you use
                  \string\begfig\space instead of \string\begfigside? }%
      \fi
      \dimen0=\fullhsize
      \advance\dimen0 by-#2true cm
      \advance\dimen0 by-1true cc\relax
      \bgroup
         \ifdim\dimen0<8true cc\relax
            \message{|No sufficient room for the legend;
                     using \string\begfigwid. }%
            \begfigwid #1cm#3\endfig
         \else
            \ifdim\dimen0<10true cc\relax
               \message{|Room for legend to narrow;
                        legend will be set raggedright. }%
               \rightskip=0pt plus 2cm\relax
            \fi
            \setbox0=\vbox{\def\figure##1##2{\vbox{\hsize=\dimen0\relax
                           \@startins\noindent\petit{\bf
                           Fig.\ts##1\unskip.\ }\ignorespaces##2\par}}%
                           #3\unskip}%
            \ifdim#1true cm<\ht0\relax
               \message{|Text of legend higher than figure; using
                        \string\begfig. }%
               \begfigwid #1cm#3\endfig
            \else
               \def\figure##1##2{\vbox{\hsize=\dimen0\relax
                                       \@startins\noindent\petit{\bf
                                       Fig.\ts##1\unskip.\
                                       }\ignorespaces##2\par}}%
               \bothtopinsert\line{\vbox{\hsize=#2true cm\relax
               \@startins\rahmen{#1}}\hss#3\unskip}\figfuzz\endbothinsert
            \fi
         \fi
      \egroup
   \fi\ignorespaces}
\def\begfigpage#1cm#2\endfig{\specialpage{\@startins
   \vskip3.7pt\rahmen{#1}#2}\ignorespaces}%
\def\begtab#1cm#2\endtab{\single@column@insert{#2\rahmen{#1}}\ignorespaces}
\let\begtabempty=\begtab
\def\begtabfull#1\endtab{\single@column@insert{#1}\ignorespaces}
\def\begtabemptywid#1cm#2\endtab{\relax
   \if N\lr
      {\hsize=\fullhsize \begtabempty#1cm#2\endtab}%
   \else
      \bothtopinsert\line{\vbox{\hsize=\fullhsize
      #2\rahmen{#1}}\hss}\medskip\endbothinsert
   \fi\ignorespaces}
\def\begtabfullwid#1\endtab{\relax
   \if N\lr
      {\hsize=\fullhsize \begtabfull#1\endtab}%
   \else
      \bothtopinsert\line{\vbox{\hsize=\fullhsize
      \noindent#1}\hss}\medskip\endbothinsert
   \fi\ignorespaces}
\def\begtabpage#1\endtab{\specialpage{#1}\ignorespaces}
\catcode`\@=\active   
%
\input aa.cmm
\font\pr=cmr7

\def\arcs{$^{\prime\prime}$}

\def\kms{km s$^{-1}$}

\MAINTITLE={  AG Carinae III.}
\SUBTITLE { The 1990 hot phase of the star and
the physical structure of the circumstellar environment$^*$}
\FOOTNOTE{ Based on observations made at the European Southern Observatory,
La Silla, Chile. }

\AUTHOR={ R. Viotti@1, V.F. Polcaro@1, C. Rossi@2}

\INSTITUTE={
@1 Istituto di Astrofisica Spaziale, CNR,
Via Enrico Fermi 21, I-00044 Frascati RM, Italy
@2 Istituto Astronomico, Universit\`a La Sapienza,
Via Lancisi 29, I-00161 Roma, Italy  }
\KEYWORDS={ stars: cirumstellar matter -- stars: emission line --
stars: individual (AG Car) -- stars: mass loss }
\OFFPRINTS={ R. Viotti }
\DATE={ received; accepted }

\THESAURUS={ 19.25.1; 19.33.1; 19.42.1; 19.48.1 }

\ABSTRACT={We report new long slit blue and red spectra of the region around
the Luminous Blue Variable AG Car obtained at ESO in June 1990.
The spectroscopic observations of the central star, observed just before
the onset of its
new brightening phase, indicates that AG Car was in the hottest
phase so far recorded. The spectrum shows strong Balmer emission lines,
and broad emissions of He {\pr II} $\lambda$468.6, and
[Fe {\pr III}] $\lambda$$\lambda$465.8-470.1,
as well as many lines of He {\pr I}, N {\pr II}, Si {\pr III}, Al {\pr III}
and Fe  {\pr III} with a P Cygni profile.
The [N {\pr II}] red doublet is present in the wings of H$\alpha$.
We have also identified weak emission lines of C {\pr II}
and, for the first time, of O {\pr II} with a P Cygni profile.

The nebula was observed with the slit centered on the star, and
displaced by $\pm$5 and $\pm$10 arcsec in declination.
All fluxes of the nebular lines (H {\pr I}, [N {\pr II}], [S {\pr II}])
peak in the ring.
The mean electron density is n$_e$= $\sim$580 cm$^{-3}$.
We also find that the spectrum of some regions 5\arcs North
and South of AG Car is partially due to scattered star's light.
This provides further evidence of the presence of dust grains
close to the central star, and indicates that even in
present times dust is condensing from AG Car wind.
A faint nearly uniform continuum emission has been detected
for the first time within the whole ring nebula which
we again associate with the presence of dust.
The existence of an effective dust condensation
process in the wind of AG Car
should largely affect the chemical abundance of the nebula.
The sharp outer edge of the ring might suggest the presence of a shock front.

We have finally identified an extended low density (n$_e$= $\sim$180 cm$^{-3}$)
H {\pr II} $halo$, which should be
associated with the residual of the stellar wind of a previous, probably
cooler, evolutionary phase of AG Car.
In the halo the H$\alpha$/[N {\pr II}] ratio is strengthened
with respect to the ring nebula,
possibly because of a discontinuity in the N abundance.
The relative line strengths in the ring and halo are those typical of large PNs
and H {\pr II} regions, respectively.
We suggest that part of the star's reddening originates in the
circumstellar ring nebula and diffuse halo, which could substantially
alter the present estimates of the star's distance.
}

\maketitle

\titlea{Introduction}

AG  Car (=HD 94910), together with $\eta$ Car and P Cyg, is  one  of the
few well established galactic Luminous Blue Variables (LBVs). These
objects  are believed to be a short-living  (10$^4$-10$^5$ years)  phase
of  the high mass  (M$\ge$40  M$_{\odot}$)  stars  evolution, immediatly
preceding the WR phase.
In spite of its many peculiarities, such as the prominent emission line
spectrum, the ample light variations, and the presence of a ring nebula,
AG Car has been only in most recent times the subject of systematic studies.
The star shows a large variability (between 6$^{m}$ and 9$^{m}$) with time
scales of the order of many years, and with
superimposed  erratic variability  of around 0.5$^{m}$ (eg. Nota et al 1992
and references therein). Caputo and Viotti
(1970, Paper I)  found that  the  luminosity variations of the star are
accompained
by dramatic spectral changes, from equivalent spectral type A1 I to
B0 I, being the star hotter near its visual light minima.
During 1981-1985 AG Car underwent a large luminosity fading in the visual,
which was accompained by conspicuous spectral changes, both at
visual and ultraviolet wavelengths (Wolf and Stahl 1982, Viotti et al. 1984,
Stahl 1987).
Viotti et al. (1984) showed that the visual fading after 1981 was
accompained by a corresponding increase in the UV flux, so that the total
bolometric  luminosity remained constant. This is an important result which
was later found in other LBVs.
Stahl (1986) found that the spectral type of the star during the January
1985  minimum was similar to that of intermediate  Ofpe/WN9 stars.
A new brightening phase started in 1990 with an increase of the visual
and IR luminosity (Mattei 1992, Whitelock 1992), and marked spectral
variations (Leitherer et al. 1992).
In  both  minimum and maximum luminosity states, the star
is characterized by a rich emission line spectrum.
Variable P Cygni profiles have been detected in different epochs in
Hydrogen Balmer, He {\pr I}, and Fe {\pr II} lines.
This fact,  as  well as the presence of  many forbidden lines  in  the spectrum
of the star is an indicator of an
extremely complex and variable structure of the stellar envelope.

A  key characteristics of AG Car is its surrounding  ring  nebula which  has
revealed  in recent times many peculiarities.
This nebula, named 289--0$^{\circ}$.1 in the Perek and Koutek (1967)
catalog  of planetary nebulae, was probably ejected by AG Car (e.g.
Thackeray 1977, Stahl 1987, Viotti et al. 1988, Paper II).
Its  UV  spectrum  is  similar  to  that  of  the  central  star,
suggesting  that  dust  grains are scattering  the  star's  light (Viotti
et  al. 1988); this hypothesis was also supported by the far-IR KAO
observations
(McGregor et al. 1988b). Paresce and  Nota (1989) discovered an  inner
dusty jet-like feature with  a  possible  helical structure. Smith (1991) and
Nota et al. (1992) studied the  nebula dynamics and found that the radial
velocities are consistent with a hollow expanding asymmetrical shell.

Thus, the studies performed up to now on AG  Car have  supplied a
large amount of experimental evidences that  can be used to clarify the LBV
evolutionary phase of very massive stars.  On the other hand, some crucial
points
are still obscure and need to be  clarified  in order to model the physical
conditions  of  the object.  For  istance,  the distance of AG Car  is
still  poorly determined.  Up  to  a few years ago, the star  was  believed
to belong  to  the  Carina OB associations and  its  distance  was  thus
estimated to be about 2.5 kpc (Viotti 1971) the same as that of the
Trumpler 14 and 16 associations (e.g. Tapia et al. 1988).
Recently, Humphreys et al. (1989) and Hoekzema et al. (1992)
suggested a  much higher value (5 to 8 kpc) on the basis of
more recent kinematic  and luminosity  data of field stars.
According to this value AG Car should have an absolute bolometric
luminosity of about --10$^{m}$, in good agreement with that of the
other LBVs near the Humpreys and Davidson (1979) instability track.
This would also imply that the evolutionary track of the star
should have never reached the red supergiant phase.

Another problem that is far to be solved is the origin and thermal balance of
the dust grains which are present in the nebula. Far-IR  data indicate that
the dust grains should be cool (60 K) and composed  by large
($\sim$1 $\mu$m) grains (McGregor et al. 1988b), in order to survive so close
to a very luminous hot star.
This fact  suggests that  the dust nebula must be old, possibly a relict of a
previous cooler evolutionary phase of the central star,
in order  to  have enough  time for large grains be formed. On the other
hand,  the colours  of the 'jet' strongly support the hypothesis that it
is mainly composed of dust grains (Paresce and Nota 1989, Nota et  al. 1992).
Nota et al. (1992) have  analyzed the  possible formation of the dust
as due to the presence of an equatorial disk due to the stellar rotation
or to a close, undetected companion.
Neither hypotheses are fully convincing  because of the  lack of other
detectable effects generally associated with these models, such as line
splitting, short term periodicities, and hard-X ray emission.
We  remind  that  the unexplained
dust formation around very hot objects (like WR  stars) is not an unusual
problem (Conti and Underhill 1988), and that only in a few cases this effect
can be explained by the presence  of  a companion (Williams and van der
Hucht 1991). In  most  of  the reported cases, these dusty nebulosities
are asymmetric.

A  further unsolved problem concerns the chemical
composition of AG  Car  and of its  nebula.  The  evolutionary models (e.g.
Maeder 1990) indicate that a LBV should be overabundant in  nitrogen.
In fact, a N/C overabundance in the star spectrum could be suggested by the
strength of the N {\pr II} lines during the $B_{eq}$ phase and the
contemporaneous undetectability of C {\pr II} and O {\pr II} lines
(Caputo and Viotti 1970). In the IR, no molecular emission from CO
was found by McGregor et al. (1988a), whereas it was detected in the
similar object HR Car.
Mitra and Dufour (1990), from the study of the spectrum of the nebula derived a
high N/CO ratio in the nebula which they attribute
to a marked CO underabundance, while the N should have a normal abundance.
Whereas, Pacheco et al. (1992)
found that nitrogen actually is overabundant by one order of magnitude
and oxygen deficient by at least a factor 6.
Nota et al. (1992) argued that the abundance anomaly in the nebula
can  be explained if the outburst that
originated  the nebula have occurred  at the beginning of the LBV  phase,
when  the chemical composition was not too changed with respect to the
previous  phase. This  explanation is supported by
the similarity of the figures
of the dynamical age of the nebula (10$^4$-10$^5$ y, Viotti et al. 1988;
Smith 1991) and of the LBV-phase  duration  (10$^4$ y,  Humphreys 1989).
However, it is difficult to explain why in one LBV (AG Car) the main
outburst occurred at the  beginning of the phase, while in the others
($\eta$ Car and P Cyg) it seems to have occurred at a much later stage.

In this paper we report new long slit spectroscopic observations of
AG Car and its nebula, made when the star was at
a crucial phase of its light history.
\titlea {Observations}
We have observed the blue (433-487 nm) and red (620-727 nm) spectrum of AG
Car and of its surrounding nebula in June 1990,
just before the onset of the recent new brightening phase of the star
(Whitelock 1992).
Two dimensional spectra were obtained at the Casse\-grain focus of the
1.52 m ESO telescope equipped with the Boller \& Chivens spectrograph and
a GEC coated CCD 576x370 pixel detector.
A 850  grooves mm$^{-1}$ grating was used, giving a resolving power of
0.18 and 0.36 nm (FWHM) in the blue and red ranges, respectively.
Excellent seeing (better than  1 arcsec) during the whole observing run
allowed the use of a 1 arcsec slit width.
The slit was 3.0 arcmin long, oriented  in the E-W direction.
Spectral images were taken with the slit
centred on the central star, and 5 and 10 arcsec to the North and to the
South. The slit positions were accurate in declination
within $\pm$1 arcsec, as also checked {\it a posteriori} from the
presence in the images of nearby stars
which were identified from the direct images.
The standard stars Wolf 485A and Kopf 27 were used for flux calibration.
Table 1 reports the detailed log of the observations.

%
The columns of the CCD images were binned during the reading process, and the
final data are recorded as
576x103  images with a spatial scale of 1.0 arcsec per pixel
in the dispersion direction, and 2.0 arcsec per binned pixel
orthogonally to it. The whole 3 arcmin slit length
corresponds to 90 binned columns.
In the dispersion direction one pixel corresponds to 0.09 and 0.18 nm
in the blue and red spectral images, respectively.
The effective spectral resolution is of 2.0 pixels.

The images were reduced with the VAX 8550 of the Istituto di Astrofisica
Spaziale
using standard procedures for background subtraction and flat-field correction.
Then one-dimensional spectrograms were extracted by adding up three adjacent
columns -- corresponding to an effective projected area of 6x1 arcsec$^2$ --
centered on the odd columns for a total of 45 spectrograms for
each frame.

\begfig 8.0cm
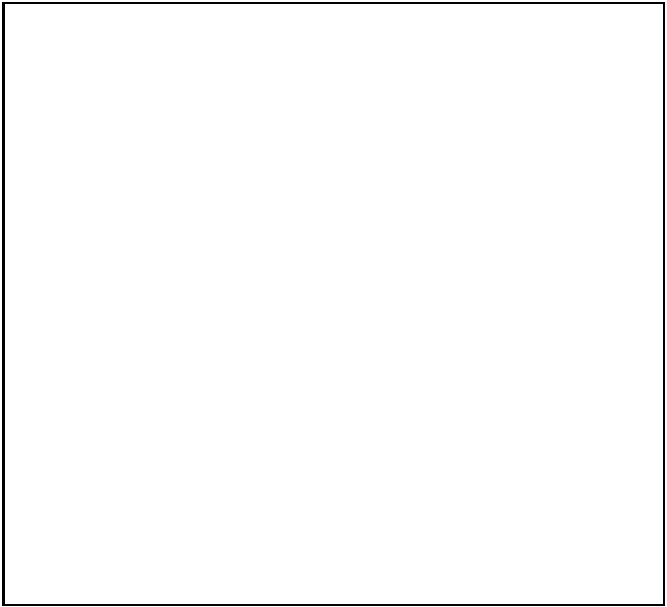
\figure{1}  { Example of the red spectrum far from the ring nebula of AG
Car. Top: the original spectrum; middle: the 'sky' spectrum; bottom: the
sky-subtracted spectrum. The original and 'sky' spectra have been
vertically shifted of 1.5 and 0.5 respectively.
Ordinates are monochromatic fluxes in 10$^{-17}$ W m$^{-2}$ nm$^{-1}$
per 6 arsec$^2$.
The peak of H$\alpha$ and
[N {\pr II}] 658.4 nm in the sky-subtracted spectrum have been cut
in order to avoid overlapping of the tracings. }
\endfig

In order to take into account the small distortion along the detector,
the wavelength calibration were performed using a polynomial approximation
on each extracted spectrogram of the He-Ar comparison lamp.
This pixel--to--wavelength
relationship was applied to the corresponding columns of the
object spectrograms.
After wavelength calibration,
the extracted spectrograms were corrected for the
atmospheric extinction and for the response curve of the night
in order to obtain absolute fluxes.
The major problem of the reduction procedure was represented by the
sky subtraction in the nebular spectra.
For extended sources the sky subtraction is often difficult, especially in
the red, as sky and nebular emissions are mixed up.
In our case many intense sky lines are visible in the 10 min red
exposures. Since the nebular emission lines are
still visible at the edges of the field,
we could not use these regions for the sky subtraction.
This is illustrated in figure 1 where
the red spectrum of the AG Car nebula near the field edges,
and the sky spectrum in the field of one
standard star, are compared.
This comparison also shows that the underlying continuum
in the AG Car nebular spectrum is mostly telluric in origin.
The accurate subtraction of this effect has enabled us to find out the true
nebular continuum near the star.
Note that the presence of sky emission lines around H$\alpha$ could be a
cause of error especially in the extracted spectrograms far from the
centre of the nebula, where the nebular emissions become quite weak.
This was found particularly worrysome for the [N {\pr II}] lines.

Having used the same exposure time for the AG Car nebular spectra
and for one of the standard stars, and thanks to the eccellent sky conditions,
the sky line and continuum emission resulted to
have the same intensity in the two sets of data.
Therefore the following sky subtraction procedure was performed:
for both the standard star and nebular frames
we have verified that the intensity of the sky lines was
constant along the frame columns; then we have obtained a mean
sky spectrum of the standard star by averaging 10 spectrograms (5 for each
side of the star); finally we have subtracted this spectrum from
the extracted spectrum of the AG Car
nebula. As shown in figure 1, the result seems to be  quite satisfying.
No local sky lines are visible in the less exposed red spectra centred on
the star and in the blue images. Thus, no sky subtraction has been performed
for these spectrograms.

{}From the cross-check  of the flux calibration obtained using the two standard
stars we are confident of an uncertainty of 10\%
for the absolute fluxes in the blue region, and of a somewhat larger
value in the edge regions.
In the red only one standard star has been taken exactly in the same spectral
region. However, during the observing run
we have observed other targets in the same red range of AG Car and in adjacent
regions which partially overlap with the red one (e.g. Polcaro et al. 1992).
Since we have found that the calibrated spectra are in good agreement in the
overlapping regions, we are confident that the accuracy for the red spectrum of
AG Car and its nebula should be comparable to that above derived
for the blue region.

In the red and blue spectrograms the noise has been measured to be
1.3 10$^{-19}$ W m$^{-2}$ nm$^{-1}$ ptp in the nebular spectra.
On the star the S/N ratio is 45 in the red continuum at 630 nm and
24 in the blue continuum at 440 nm. Referring to the lines the S/N ratio
obviously depends on the line intensity; for instance we find S/N=132
at the H$\alpha$ line peak, and 50 at the He {\pr I} $\lambda$447.1 line peak.

\begfig 17.0cm
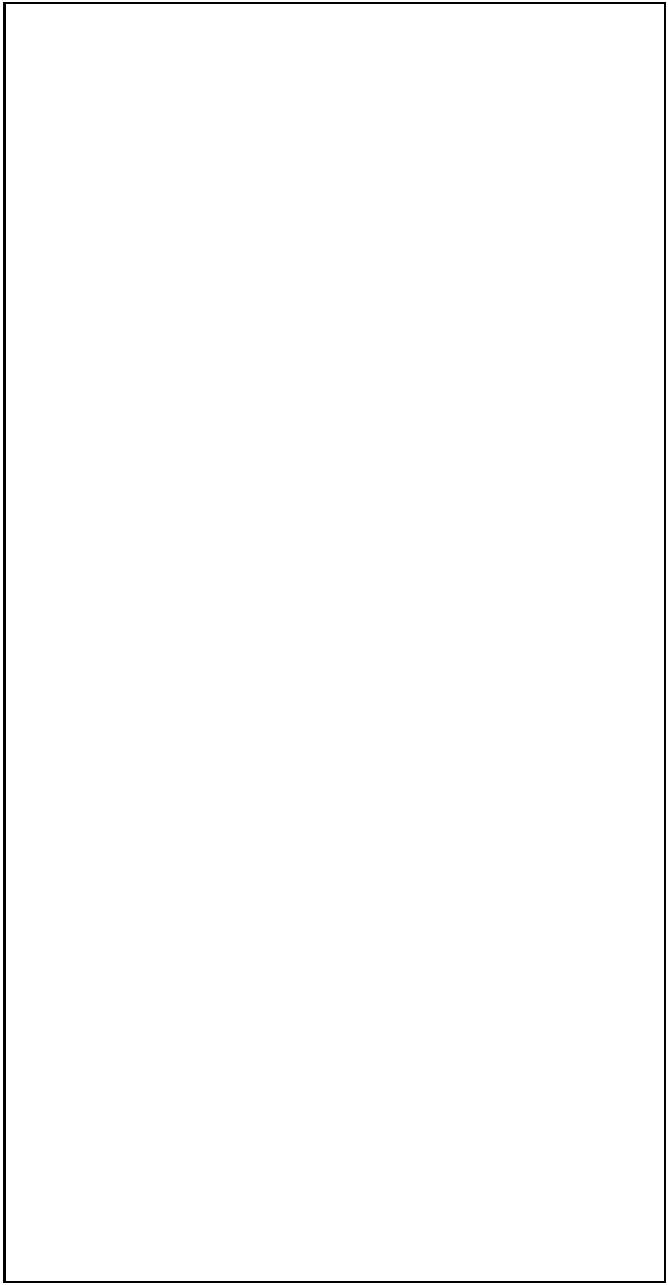
\figure{2}  { (a) The blue spectrum of AG Car in June 1990.
(b) The red spectrum of AG Car observed in June 1990.
(c) Enlargement of the blue spectrum to show the different profiles of the
He {\pr I}, He {\pr II}, N {\pr II} and [Fe {\pr III}] lines.
Ordinates are monochromatic fluxes in 10$^{-14}$ W m$^{-2}$ nm$^{-1}$,
not corrected for the interstellar extinction.
}
\endfig

\titlea {Analysis of the results }
\titleb {The star }
The spectrum of AG Car was measured on the central
extracted spectrogram  of the CCD frames centred on the star.
The calibrated spectra in the blue and red regions are
shown in Figure 2.
Line fluxes were measured following the procedure described in
previous articles (e.g. Polcaro et al. 1992, Viotti et al. 1989), and the
results are reported in Table 2 where the successive columns give:
(1) Heliocentric line barycentre (in nm).
(2) Type of the line: absorption (A) or emission (E).
(3) Total width of the line.
(4) Line equivalent width (in nm); "p" = line present but not
measurable.
(5) Adopted continuum level in 10$^{-16}$ W m$^{-2}$ nm$^{-1}$
(i.e. 10$^{-14}$ erg cm$^{-2}$ s$^{-1}$ \AA$^{-1}$) local to the feature,
not corrected for the interstellar absorption.
(6) Identification of the ion contributing to the observed line;
(7) Laboratory wavelength.
(8) Remarks: "PCy" = P Cygni violet-shifted absorption; "i.s." = interstellar
line; "bl" = blended with the nearby line; "br" = broad line;
possible contributors are also indicated.

Our observations were performed in an important phase of the star light
curve since at the time of our observations AG Car was near
minimum luminosity (Mattei 1992),
and a few days later it started to brighten
again, both in the visual and in the IR (Mattei 1992,
Leitherer et al. 1992, Whitelock 1992).
Our spectrum is characterized by the presence of very
intense H$\alpha$, H$\beta$ and H$\gamma$ emission lines,
strong He {\pr I} lines with P Cygni profile, and of
emission lines of He {\pr II} $\lambda$ 468.6$\;$nm
and [Fe {\pr III}] $\lambda\lambda$ 465.8 and 470.1$\;$nm
definitely broader than the instrumental profile.
In addition, we have identified P Cygni lines of Si {\pr III},
Al {\pr III}, Fe {\pr III}, and especially of N {\pr II} (Fig.2).
The optical spectrum of AG Car also shows weak
lines of O {\pr II} $\lambda$464.9,
Si {\pr II} $\lambda$637.1, and
C {\pr II} $\lambda\lambda$723.1-723.6.
The underlying continuum is rather flat, in agreement with the fact that
AG Car is a highly reddened blue star.
The B and V magnitudes, as estimated from the continuum energy distribution,
are $\sim$9.1 and $\sim$8.6, respectively.
Whereas the colour index is in agreement
with the stellar colour excess reported in the literature,
the visual magnitude is a few tenth of magnitude fainter than that reported by
AAVSO for the same period. This discrepancy could be partly attributed to the
calibration error, and partly to a not negligible
contribution of the emission line flux to the broad band photometry.
We cannot anyhow rule out the possibility that AG Car was observed
during a deep visual luminosity minimum.

No blue-shifted absorption component has been observed in the
Balmer lines. Actually, the disappearence of the P Cygni absorption
in H$\alpha$ during the hotter phase of AG Car was already noted by Bandiera et
al. (1989), and should be associated with the increased ionization
of hydrogen.
Since H$\alpha$ is saturated also in the lower exposed red spectrum,
we cannot make an estimate of its intensity.
The H$\gamma$/H$\beta$ flux ratio of 0.37 (uncorrected for the interstellar
extinction) is very similar to that measured in the ring nebula as discussed
in the next section. This value is also in agreement with the value for
the 1984 spectrum of AG Car found by Mitra and Dufour (1990),
in spite of the different photometric phase of the star.
This fact not only confirms that the contribution
of the dust to the extinction in the optical wavelengths is small (McGregor
et al 1988b, Mitra and Dufour 1990), but also that it is not variable with the
time or with the photometric stage of the star.

The He {\pr II} line presents a remarkable round-topped shape with a FWHM,
corrected for the instrumental profile, of about 220 \kms (Fig.2c). Its
strength indicates that at the time of our observations AG Car was in a
somewhat hotter phase than that observed in 1985 (Stahl, 1986).
The high temperature phase is also marked by the large number of emission
lines of highly excited species.
The N {\pr III} 463.4-464.2 nm multiplet, which is characteristic of Of stars,
if present, is masked by the many strong N {\pr II} lines (Fig.2c).
This ion was not present in the Stahl's 1985 spectrogram.
The [N {\pr II}] doublet is present in the wings of H$\alpha$ (Fig.2b).

A rich N {\pr II} emission spectrum was previously observed in the
spectrum of AG Car during the high excitation phase of 1953-1955
(Caputo and Viotti 1970), but never simultaneously to
the He {\pr II} 468.6 nm line. Thus the 1985-1990 phase
was the hottest one so far observed in AG Car.
We have also for the first time identified O {\pr II} in the spectrum
of AG Car (Fig.2c), while
the C {\pr II} $\lambda\lambda$723.1-723.6$\;$nm
doublet is present in the red spectrum; the stronger C {\pr II}
$\lambda\lambda$
657.8-8.3$\;$nm doublet is probably hidden in the intense
H$\alpha$-[N {\pr II}] blend. These latter
lines have been identified in the very high resolution CAT-CES observations
of February 1984, during the fading phase of AG Car
when the [N {\pr II}] lines were absent (Bandiera et al. 1989).
The C {\pr II} $\lambda$ 657.8 line was still visible in the June 1987
spectrum in the wings  of the [N {\pr II}] 658.4 nm line (Bandiera et al.
1989).
The O and C ions were not identified in the earlier spectra of
Caputo and Viotti (1970), but this could be at least partially attributed to
the
lower quality of their spectra.

The richness in nitrogen lines of the spectrum of AG Car has
suggested previous workers to hypothize a N overabundance in
AG Car, in agreement with some theoretical expectations. On the other hand
the complex structure of the atmospheric envelope of this star
might largely enhance some spectral lines and depress others,
so that a reliable abundance determination is impossible without
a good knowledge of the wind structure (see for instance the
case of the carbon abundance in $\eta$ Car discussed by Viotti et
al. 1989). However, as discussed in the following section,
a large N/O overabundance
has been firmly found in the circumstellar AG Car nebulosity.

Another important argument is the velocity field
in the wind of AG Car. In particular we have found that the high
excitation He {\pr II} and [Fe {\pr III}] emission lines appear
broad, suggesting line formation in an expanding medium
with a velocity of about 200 \kms. A similar velocity broadening is
measured for the [N {\pr II}] lines in the 1987 spectrum
taken with the ESO CAT-CES (Bandiera et al 1989).
Indeed, a careful analysis of the profile of the [N {\pr II}] 658.4 nm
line in our, lower resolution red spectra confirms that these lines are
broader than the other intermediate excitation wind features.

Most of the permitted lines in the 1990 spectrum of AG Car have
a P Cygni profile with an E-A velocity difference from $\sim$150 to
$>$200 \kms, comparable with the broadening velocity measured
in the above discussed high excitation emission lines.
Our resolution of about 120 \kms does not allow us to look for systematic
velocity difference between different lines. We can only
argue that the terminal wind velocity in mid 1990 should
equal to or larger than $\sim$250 \kms.

It may be interesting to compare the radial velocities
meaasured in different phases of AG Car.
Caputo and Viotti (1970) found the mean E-A velocities to be about
100 \kms and 160 \kms for the H and He lines respectively.
Wolf and Stahl (1982) described a 12 A mm$^{-1}$ blue spectrum of
AG Car observed in December 1981
near its maximum visual luminosity. This spectrum is characterized by
many Balmer and Fe {\pr II} lines with a P Cygni profile.
The absorption components have a minimum at about 100-110 \kms,
and their wings extend to about 150 \kms.
In an unpublished 12 A mm$^{-1}$ spectrum of AG Car
obtained by Altamore and Viotti
at the Coud\'e focus of the 1.52 m ESO telescope in February 1983,
when the star was one magnitude below maximum,
the P Cygni absorptions of the strongest Balmer and Fe {\pr II} lines
appear double, with minima at E-A=90-100 and $\sim$180 \kms.
The wings are slightly more extended than in 1981, up to about
180 \kms.
\begfig 16.0cm
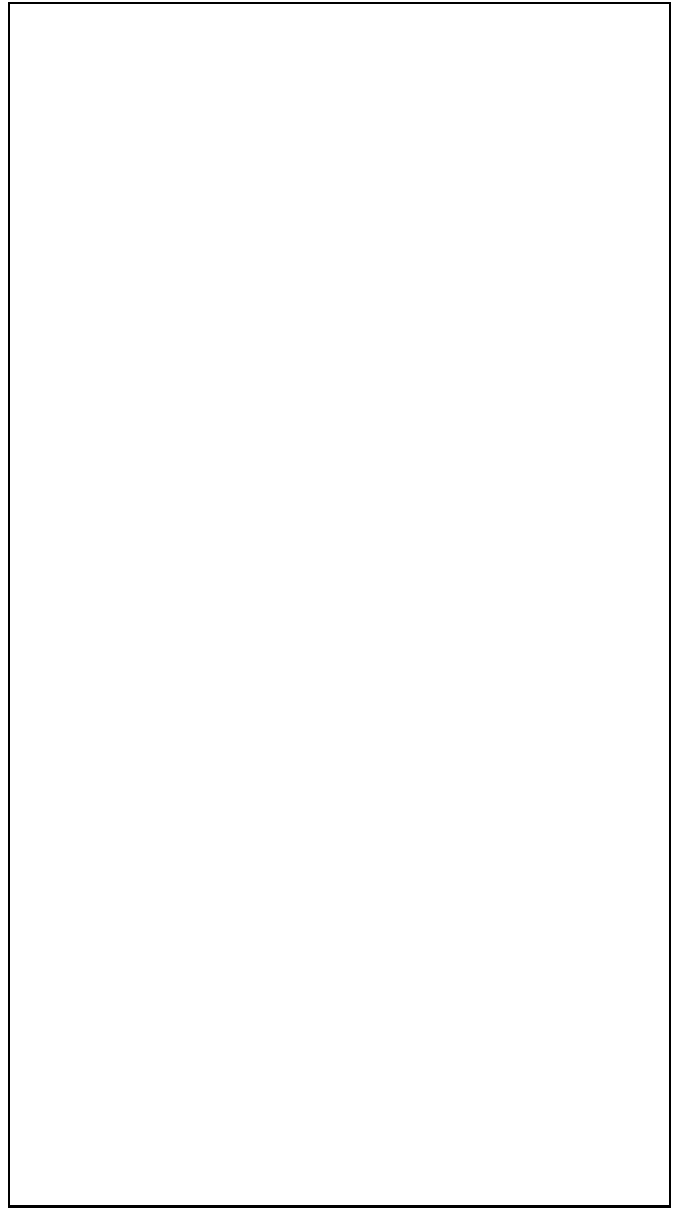
\figure{3}  { The red spectrum of the nebula 5\arcs N (a) and 10\arcs N,36\
arcs E (b).
Note the the presence of He {\pr I} lines with P Cygni profile
in the 5\arcs N spectrum.
Ordinates are monochromatic fluxes in 10$^{-17}$ W m$^{-2}$ nm$^{-1}$
per 6 arsec$^2$.}
\endfig
\titleb{ The emission line spectrum of the nebula }
As discussed in Section 2, from each CCD frame we have extracted a set of
45 calibrated spectrograms which have been used to derive the line and
continuum fluxes across the ring nebula and outside it.
The continuum data will be discussed in the next section. As already said,
the first inspection of the spectra extracted from the "nebular" frames
has revealed that the nebular emissions are detectable
still far outside the ring.
This fact indicates that the ring nebula is embedded in a diffuse nebular
region (or $halo$). The presence of this halo is confirmed by
an unpublished CCD H$\alpha$
image of AG Car kindly put at our disposal by A. Altamore.
Both our spectroscopic observations and the H$\alpha$ image
show that the nebular emission decreases outwards, indicating that it
is associated with the star, rather than with the general Carina arm H {\pr II}
emission.

We have measured the flux of the following lines:
H$\alpha$, H$\beta$, H$\gamma$, [N {\pr II}], and [S {\pr II}].
The nebular [O {\pr I}] 630-637 nm emission could also be present,
but the doublet is largely masked by the [O {\pr I}] sky lines.
For the CCD frames centered on the star the smaller esposure times
prevented us to derive good line fluxes much outside the central star.
Two blue frames (5\arcs N and 10\arcs S) appear not well exposed in spite of
the fact
that the exposure time was the same as for the other images. Although the
H$\gamma$/H$\beta$ flux ratio is in perfect agreement with those of the other
frames, they have not been used for this study.
Some spectra of the nebula in the red region are shown in Figure 3.
The central spectra of the frames centered 5\arcs~South and North present a
large similarity with the stellar one, suggesting that the nebular spectrum
is  partially due to starlight scattered by dust grains.
This point will be disscussed in the following section.

\begfig 16.0cm
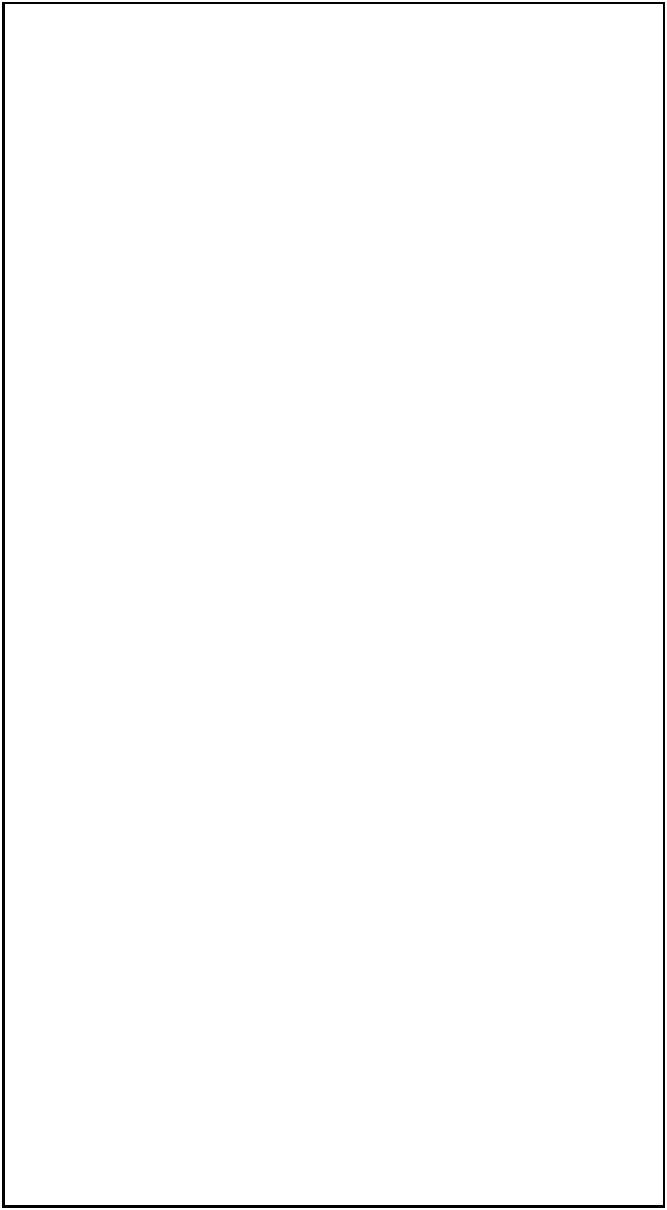
\figure{4}  { The integrated flux across the nebula at different slit
positions for the main emission lines.
Ordinates are fluxes in 10$^{-17}$ W m$^{-2}$.
East is to the right. The blue 5\arcs N and 10\arcs S scans are of lower
quality. }
\endfig

The measurement of the line fluxes has been performed by integrating
the line profile between fixed couples of wavelength points, and using
the same regions for the continuum level measurement.
The results are plotted in Figure 4 where we report the line fluxes measured
across the nebula.
In the inner positions (0\arcs, 5\arcs N and 5\arcs S) two
maxima are visible on the opposite sides of AG Car which correspond to the
two sides of the ring nebula crossed by the slit. In the 10\arcs N and 10\
arcs S
frames the two maxima are blended into a single peak, as expected from the
ring geometry.
For the [S {\pr II}] lines the central points have been omitted in the
0\arcs~images, as these lines are absent in the stellar spectrum.

Figure 4 indicates that for all the lines the position and
the relative intensity of the two peaks changes in a continuous way with the
declination, showing that both the geometry and the emission column
density largely deviate from spherical symmetry.
Note also that a large nebular emission is present inside the ring.
The spatial distribution of the intensities does not seem to be correlated with
the jet like structure discovered by Paresce and Nota (1989).

\begfig 11.0cm
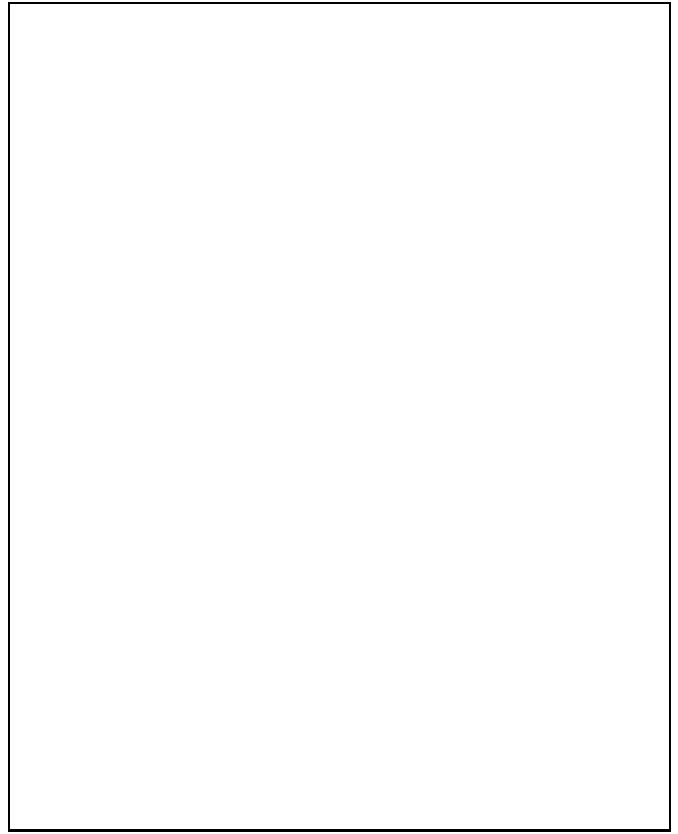
\figure{5}  { Flux ratio of the main lines, not corrected for the interstellar
absorption. The open squares correspond to data from the nebular frames;
the triangles correspond to the data from the frame centered on the star.
(a) H$\gamma$/H$\beta$; (b) [S {\pr II}] 671.7/673.1 nm;
(c) H$\alpha$/[S {\pr II}]; (d) H$\alpha$/[N {\pr II}].
In panels c and d the central higher points corresponds to enhanced H$\alpha$
emission due to the scattered star light.}
\endfig

Information about the physical conditions across and outside the ring nebula
can be derived from the emission line ratios which are plotted in
Figure 5.
Concerning the Balmer decrement (Fig. 5a) we have computed the
H$\gamma$/H$\beta$ ratio for all the slit positions inside the ring nebula
only, because the intensity of H$\gamma$ vanishes immediately outside the ring.
By averaging all the ratios along the slit we obtained a mean value of
0.36$\pm$0.12(rms). When dereddened with E$_{B-V}$=0.7,
this figure corresponds to 0.47 which is in agreement with
the theoretical case B (T$_e$=10000 K, n$_e$=100 cm$^{-3}$) value.
However, the ratio seems to be variable across the nebula, being equal
to 0.33$\pm$0.05, in the central parts of the nebula,
and to 0.40$\pm$0.11 in the rims. This fact could be related to
a larger extintion inside the ring nebula, but needs to be confirmed
by new higher S/N observations.

For the H$\alpha$/H$\beta$ flux ratio only the frames 10\arcs N and 5\arcs S
have
been used. The ratio show a large scatter which is probably due to the
fact that the lines are measured in different frames.
Inside the ring nebula the average value is 5.73$\pm$0.3
in agreement with the value recently found by Pacheco et al (1992).
In the log(H$\gamma$/H$\beta$) vs. log(H$\alpha$/H$\beta$) plane (Cox and
Mathews 1969) the representative point of the AG Car nebula is
placed near the reddening line for pure recombination.

The [S {\pr II}]$\lambda$671.7/673.1 flux ratio can be used to estimate of
the electron density. These lines, in spite of their weakness,
are measurable till the end of the slit in all the nebular spectra
and the plot of the line ratio for all the spectral regions but
excluding the region near the star,
is given in Figure 5b. The mean ratio inside the nebula and outside it,
is 0.95$\pm$0.03 and 1.20$\pm$0.08, respectively. The corresponding electron
density within the nebula (for T$_e$=10000 K) is
n$_e$=580$\pm$60, slightly lower than the value derived by Mitra and Dufour
(1990)
and by Pacheco et al. (1992). Outside the nebula a much lower density of
n$_e$=180$\pm$80 is derived.

Figure 5c shows the ratio H$\alpha$/[S{\pr II}]$\lambda$671.7+673.1.
In the plot the inside and outside
regions of the nebula are clearly marked, with a mean value of the ratio
changing from 12.8 to 3.5.
The anomalous maximum of about 20, not included in the mean,
corresponds to the region 5\arcs N of AG Car
where the nebular spectrum is partially stellar scattered light, with strong
H$\alpha$ emission, as discussed in the next section.
Note also that except in the above case there is no large variation of the
ratio within the nebula, in spite of the large variation of the
emission line intensity. The transition from the inner to the outer
region is rather sharp, and takes 8-10 arcsec.

The H$\alpha$/[N{\pr II}]$\lambda$654.8+658.4 flux ratio is shown in Figure
5d.
The ratio is larger in the external regions (2.4$\pm$0.08) than inside
the nebula (1.37$\pm$0.06, excluding the central higher points due to
the scattered star light).
The latter is slightly lower, but still within the
errors, than the values
given by Mitra and Dufour (1990) and Pacheco et al. (1992).
It is interesting to compare the above derived ratios for the inner and
outer regions with the typical values for different astrophysical
objects, such as Planetary Nebulae, H {\pr II} regions and SNRs.
For this purpose we have used the diagnostics of Sabbadin et al.
(1977) who used the intensity ratios of H$\alpha$,
[N {\pr II}], and [S {\pr II}] lines. We have found that in all the diagrams,
671.7/673.1 vs. log(H$\alpha$/[N {\pr II}]),
log(H$\alpha$/[N {\pr II}]) vs. log(H$\alpha$/[S {\pr II}]), and
671.7/673.1 vs. log(H$\alpha$/[S {\pr II}]),
the representative points of the ring nebula
of AG Car, and of the halo
lie well within the PN and H {\pr II} regions, respectively.
This confirms that the conditions in the outer regions
are similar to those of typical diffuse nebulae, while the
ring nebula is physically similar to large PNs.
\titleb { The dust scattered nebular continuum }
{}From a visual inspection of the CCD spectroscopic frames it appears
evident that a diffuse continuum emission is present in the regions
close to centre of the ring nebula.
The possibility should be considered that it could at least partially due
to stellar light scattered inside the instrument. However, we have found
this continuum also in those frames corresponding to observations
with the slit not centered on the star.
In addition, an inspection of the frames of standard stars has
shown that no scattered light of similar strength and
extension is present. This fact  makes us confident that the continuum
observed in AG Car actually is nebular emission.
We have measured the continuum flux in two emission-line free
regions centered at 470 and 640 nm.
In order to enhance the S/N ratio, 10 successive
raws have been added up, which corresponds to an integration
over 1 and 2 nm for the blue and red region, respectively.
Figure 6 shows the variation of the blue and red continuum
across the nebula at the different slit positions.
Except for a marked anomaly for the 5\arcs N frame near the star,
the overall picture is that of a rather smoothed emission
within the nebula, without an indication of an enhanced
emission near the rings, as seen in the emission line plots (Fig.4).
This gives evidence of a diffuse continuum
emission from the whole AG Car nebula.
Previous broad-band images have failed to detect the nebula in the blue
(A. Altamore, private communication), or have only revealed the presence
of a few peculiar features in the visual (Paresce and Nota 1989),
or in the red (Nota et al. 1992).
We attribute this diffuse continuum emission to the
presence of dust grains in the whole extension of the nebula,
which scatter the light from the central star.
We have also noted that the continuum emission sharply decreases
near the outer borders of the nebula.

\begfig 11.0cm
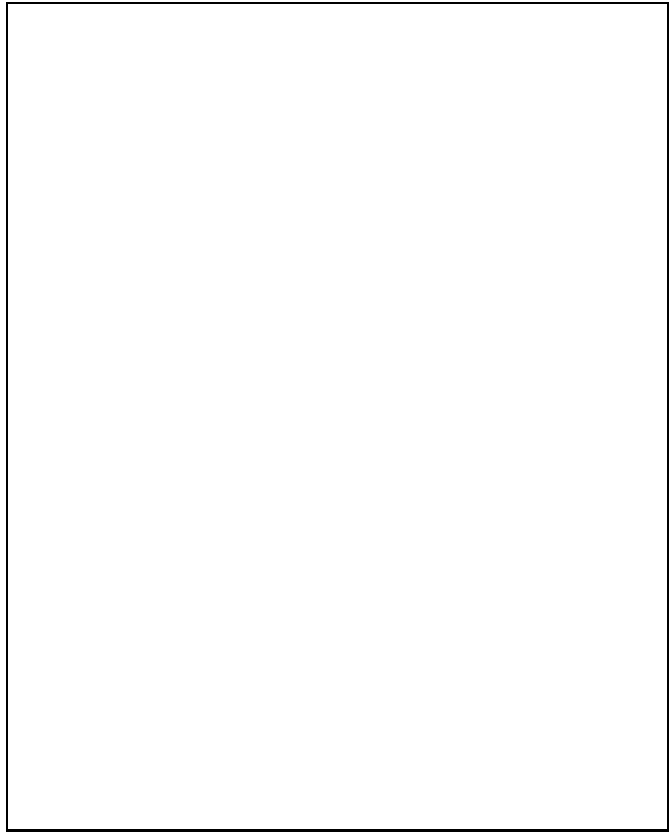
\figure{6} { The continuum across the nebula for the different
slit positions centered at 470 nm and 640 nm in the blue and in the red,
respectively. In the y axis the CCD counts corrected for bias and flat
field are reported.
Ordinates are CCD counts corrected for flat field and bias.
For the frame centered on the star, the logarithm
of the ordinates has been plotted.
Note the enhanced red continuum flux 5\arcs S and, especially, 5\arcs N of the
star.
The blue 5\arcs N and 10\arcs S scans are of lower quality. }
\endfig

In addition to the diffuse emission, we have found
that the red nebular continuum is particularly enhanced in
two regions 5\arcs N and 5\arcs S of AG Car (figures 3 and 6), extending to a
few arcsecs in
right ascention from the central star,
in agreement with Nota et al. (1992).
The spectrum 5\arcs N appears quite similar to that of the star,
with very strong H$\alpha$ and [N {\pr II}] emission
and the He {\pr I} 667.8 nm line with a marked P Cygni profile.
The interstellar 628 nm band is also clearly visible.
It is evident that the spectrum is partially
nebular, as strong [S {\pr II}] lines are present, and
partially scattered starlight.
This scattered light has to be attributed to
the presence of an increased amount of dust grains near the
star. It should be noted that this position is at the
opposite part of the dusty jet found by Paresce and
Nota (1989). Actually, also in the position of the jet
we have found emission due to scattered starlight,
but less intense than in the anti-jet direction.
\titlea {Towards a model of AG Car }
Our observations of AG Car were made at a deep photometric minimum
of the star which, in the constant bolometric luminosity model of
Viotti et al. (1984), should correspond to a minimum size of the
effective photosphere, about 10 times smaller than during the visual
luminosity maximum of 1981.
The June 1990 spectrum of AG Car present a lot of intriguing features,
which are difficult to explain within a simple  framework.
Of special interest is the presence of a broad He {\pr II}
468.6 nm emission line with an intensity much larger than that previously
measured in 1985 (Stahl 1986), and
in December 1990 (Leitherer et al. 1992) after our observations,
indicating that AG Car was in the hottest phase so far recorded,
in agreement with its low visual luminosity.
The He {\pr II} line is probably formed in the innermost layers of the
AG Car wind.
Its width indicates that in these layers the expansion velocity
is large and close to the maximum wind velocity as discussed below.

Most of the emission lines identified in the June 1990 spectrum
belong to ions within a rather small range of ionization energy
(24 to 35 eV), suggesting a high wind temperature with a rather small
temperature gradient.
All the intermediate ionization
lines display a P Cygni profile with an absorption component blue shifted
up to $>$200 \kms, suggesting a terminal wind velocity in excess
of this figure. In this regard,
it should be noted that no P Cygni absorption component was displayed by the
N {\pr II} lines in the spectra taken earlier in the minimum phase (Stahl
1986).
This fact seems to suggest an increased density of the stellar wind
in June 1990,
which can be associated with the beginning of the new brightening
phase of the star which started a few days after our observations.

Another peculiarity of the stellar spectrum is the presence of
[Fe {\pr III}] lines with a flat topped,
possibly double peaked profile, which is reminiscent of the profile
of the [N {\pr II}] lines in the high resolution 1987 spectrum
of AG Car described by Bandiera et al. (1989).

AG Car is not unique in displaying broad high excitation
forbidden lines. For instance Israelian and de Groot (1992)
have recently found that in P Cyg the [N {\pr II}] lines
are broad with a velocity width of 160 \kms.
Damineli Neto et al. (1993) showed that also in $\eta$ Car the [N {\pr II}]
lines have a complex profile with a total width of about 400 \kms.
Since [Fe {\pr III}] and [N {\pr II}] should both be formed in the less
dense outer parts of the wind of AG Car, their profile clearly indicates
that the wind is still expanding at a velocity much larger than 100 \kms
quite far from the stellar effective surface.
On the other hand, much lower motions have been observed in the nebula
a few arcseconds from AG Car (e.g. Smith 1991), which implies a slowing
down of the wind quite close to the star.

A further problem concerns the chemical composition of the stellar wind
and of the ring nebula. If AG Car is an evolved massive star,
we should expect marked composition anomalies.
As discussed above, a CO underabundance of the stellar wind could
be accounted for the weakness of the P Cygni C {\pr II} and O {\pr II}
lines with respect to those of N {\pr II},
but other effects, such as the different ionization energies should be
also taken into account for a more precise estimate. On the other
side our spectroscopic study of the nebula confirms previous conclusions
that the matter in the nebula is overabundant in nitrogen.
The point is whether this overabundance is that of the central star, or it
is at least partially the result of some physical processes in the nebula.

\begfig 8.0cm
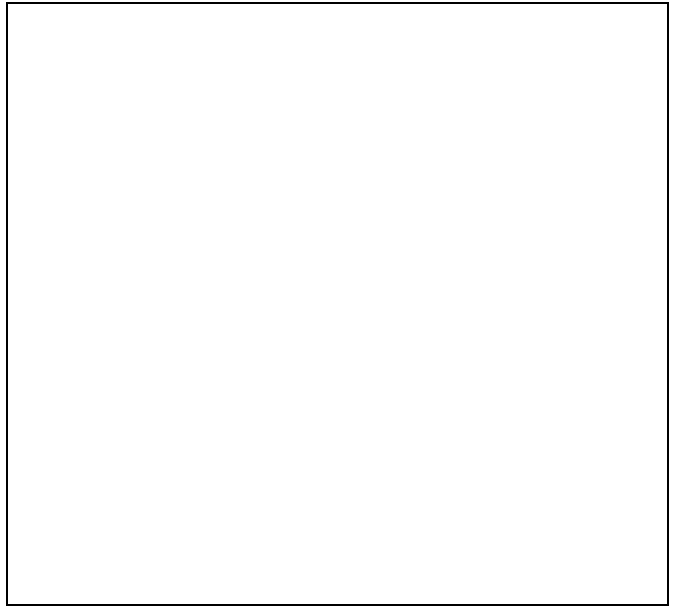
\figure{7}  { Schematic model of AG Car and its nebula,
illustrating, not in scale, the variable apparent stellar size,
the boundary of the dust condensation region, the ring and the halo.
The approximate position of some field stars is also given. }
\endfig

As discussed in section 3.3, dust signatures have been found in
different parts of the circumstellar environment of AG Car,
even in regions very close to the star.
This has led us to conclude that {\it even in present times
dust is continuously condensing from AG Car's wind }.
Since carbon and oxygen are among the main components of
different types of condensate, their possible underabundance
in the stellar wind is difficult to reconcile with the presence
of dust in the nebula, unless different types of condensate are considered.
In this regards, we should mention the absence of CO molecular emission
in the near-IR spectrum of AG Car, while it was detected in the similar
star HR Car (McGregor et al. 1988a). We also recall that dust is
continuously condensing from the $\eta$ Car's wind, in spite of its possible
carbon underabundance (Viotti et al. 1989).
Another problem concerns the physical conditions,
low temperature and especially high particle density, required for dust to
condense. This point was discussed by Andriesse et al. (1978) for the
case of $\eta$ Car.
In this star as well as in AG Car the dust condensation is probably
favoured by the breaking of the outflowing wind into denser cloudlets.
This process could be associated with the change of the expansion velocity
discussed above.
Actually, the "jet-like"
structure discovered by Paresce and Nota (1989) and the evidence (that we
have confirmed) of a high dust density in the "anti-jet" position
points to the possibility of an axisymmetric enhancement of the dust
concentration.
The jet itself could be composed by many badly resolved cloudlets.
Whatever the mechanism of dust condensation be,
it would reduce the CO abundance of the ejected gas,
and be at least partially responsible of the anomalous
abundance of the ring nebula.

Another piece of the puzzle is the structure of the nebula. Our data show
that the ring nebula is surrounded by a much less dense, apparently spherically
symmetric H {\pr II} region which can be considered a residual of the stellar
wind (or $halo$) of a previous evolutionary phase of the star. Indeed, as
discussed in section 3.2, the nitrogen abundance in this halo is probably
normal.
The measure of the expansion velocity of the halo is crucial to understand
whether it was ejected by a red supergiant. Only an upper limit of about 100
\kms can be derived from our low resolution spectrograms.

The outer boundary of the ring appears
very sharp, both from a morphological and a spectroscopic point of view,
which might suggest the presence of a physical discontinuity, i.e.
a shock front. In any case the
transition from the phase when the outer halo and the inner ring were formed
was a discontinuty in the star evolution.
\titlea {Conclusions }
Our observations have disclosed a number of important data on the
AG Car environment, which are schematically illustrated in figure 7.
{}From the He {\pr II}
profile we have argued that during the 1990 hot
phase, the stellar wind reached its velocity maximum near the stellar
surface, indicating the presence on an effective wind acceleration mechanism.
Then, at a distance of 10$^{16}$-10$^{17}$ cm, the outflowing matter is
breaking into a large number of cloudlets, with a larger
cloud density in the jet and anti-jet directions.
The unknown breaking process is probably also helping the process
of dust condensation and growth.
Since dust grains seem to be present everywhere, in the ring nebula as
well as inside it, the dust condensation should have worked for at
least 10$^4$ yr. This could be partially responsible of
the CNO anomaly in the nebula.
On the other hand, in the absence of reliable wind models, it is
at present impossible to relate the observed weakness of the C {\pr II} and
O {\pr II} P Cygni lines in the stellar spectrum with a CO underabundance.

All these observational facts cannot be explained by any simple model of the
AG Car nebula, including the presence of a close companion
or of a magnetic field as for instance suggested by Nota et al. (1992).
On the contrary, they point to a complex interaction
of different velocities in the wind, associated with different evolutionary
phases of the central object, similar to those studied by Pascoli (1992)
for the planetary nebulae formation, or to those suggested by
Wang and Mazzali (1992) for the "Napoleon hat" nebula surrounding SN1987A.
These interactions can also be the reason of the velocity discontinuity
between the wind and the inner nebula.

A last evolutionary consideration follows from the determination of the
star distance. As mentioned above, the most recent evaluation of the
AG Car distance seems to favour a value $\simeq$ 6 kpc
(Humphreys et al. 1989, Hoekzema et al. 1992).
If this figure is true, the corresponding large absolute luminosity would
imply, following the current models,
an evolutionary track not reaching the RSG phase.
However, the large reddening of AG Car (E$_{B-V}$$\sim$0.7)
could be partially attributed to an excess of local reddening,
due to absorption from circumstellar dust in the nebula as well
as in the extended halo.
An extinction of A$_V$$\sim$0.025-0.05 due to the cool dust in the
ring was derived from the KAO far-IR observations by McGregor et al. (1988b).
Also the innermost dusty regions discussed in this paper should
contribute to a local extinction.
The quantitative evaluation of the contribution of the hotter
circumstellar grains to the stellar colour excess
would require a detailed study of the IR energy distribution of the star,
which is dominated by the free-free emission from the stellar wind
(e.g. Bensammar et al. 1981).
If the AG Car distance is that of the Carina complex, as suggested by
Viotti (1971), its absolute luminosity (and thus its mass) should be
definitely lower than that of the other LBVs. In this case the star
can have experienced an RSG phase and would
be now in an evolutionary phase that has not been previously recognized.
\acknow{ This research has benefitted of the use of the SIMBAD database
operated at CDS, Strasbourg. We also thank Janet Mattei and
Patricia Whitelock for having provided us with the recent optical and
IR light curves of AG Car, and to Aldo Altamore for the CCD H$\alpha$
image of AG Car. O. Stahl is aknowledged for useful comments.
The extracted spectrograms of AG Car and of its nebula
are available as computer files on request to C.R.
at the SPAN address 40058::ROSSI. }
\begref {References}
\ref Andriesse C.D., Donn B.D., Viotti R., 1978, MNRAS 185, 771
\ref Bandiera R., Focardi P., Altamore A., Rossi C., Sthal O., 1989,
  Physics of Luminous  Blue Variables.
     In: Moffat F.G., Davidson K. (eds.) Proc. IAU Coll.113,
     Kluwer, Dordrecht, p.279
\ref Bensammar S., Gaudenzi S., Rossi C., Johnson H.M., Th\'e P.S.,
     Zuiderwijk E.J., Viotti R., 1981, Effects of Mass Loss in Stellar
Evolution.
     In: Chiosi C., Stalio R. (eds.) Proc. IAU Coll. 59,
     Reidel, Dordrecht, p.67
\ref Caputo F., Viotti R., 1970, A\&A 7, 266 (Paper I)
\ref Conti P., Underhill A., 1988, O stars and Wolf-Rayet stars,
     NASA SP-497.
\ref Cox D.P., Mathews W.G., 1969, ApJ 155, 859
\ref Damineli Neto A., Viotti R., Baratta G.B., de Araujo F.X. 1993,
     A\&A 268, 183
\ref Davidson K., Walborn N.R., Gull T.R. 1982, ApJ Letters 254, L47
\ref de Freitas Pacheco J.A., Damineli Neto A., Costa R.D.D., Viotti R.,
     1992, A\&A 266, 360
\ref Hoekzema N.M., Lamers H.J.G.L.M., Genderer A.M., 1992, A\&A 257, 118
\ref Humphreys R. 1989, Physics of Luminous  Blue Variables.
     In: Moffat F.G., Davidson K. (eds.) Proc. IAU Coll.113,
     Kluwer, Dordrecht, p.3
\ref Humphreys R.M., Davidson K., 1979, ApJ 232, 409
\ref Humphreys R.M., Lamers H.J.G.L.M., Hoekzema N., Cassatella A.,
     1989, A\&A 218, L17
\ref Israelian G., de Groot M. 1992,
     Nonisotropic and Variable Outflows from Stars.
     In: Drissen L., Leitherer C., Nota A. (eds.) Proc. STScI Workshop,
     ASP Conference Series, Vol.22, p.356
\ref Lamers H.J.G.L.M, de Groot M.J.H. 1992, A\&A, 257, 153
\ref Leitherer C., Damineli Neto A., Schmutz W., 1992,
     Nonisotropic and Variable Outflows from Stars.
     In: Drissen L., Leitherer C., Nota A. (eds.) Proc. STScI Workshop,
     ASP Conference Series, Vol.22, p.366
\ref Maeder A., 1990, A\&AS 84, 139
\ref Maeder A., Meynet G. 1987, A\&A 182, 243
\ref Mattei J.A. 1992, AAVSO observations
\ref McGregor P.J., Hyland A.R., Hillier D.J. 1988a, ApJ 324, 1071
\ref McGregor P.J., Finlayson K., Hyland A.R., et al. 1988b, ApJ 329, 874
\ref Mitra P.M., Dufour R.J. 1990, MNRAS 242, 98
\ref Nota A., Leitherer C., Clampin M., Greenfield P., Golimowski D.
     1992, ApJ, 398, 621
\ref Pascoli G. 1992, PASP 104, 350
\ref Paresce F., Nota A. 1989, ApJ 341, L83
\ref Perek  L., Kohoutek L. 1967, Catalogue of  Galactic
    Planetary Nebulae, Prague Academy. Prague
\ref Polcaro V.F., Viotti R., Rossi C., Norci L. 1992, A\&A 265, 563
\ref Sabbadin F., Minello S., Bianchini A. 1977, A\&A 60, 149
\ref Smith L.J. 1991,
     Wolf-Rayet Stars and Interrelations with Other Massive Stars in Galaxies.
     In: van der Hucht K.A., Hidayat B. (eds.)
     Proc. IAU Symp. 143, Kluwer, Dordrecht, p.385
\ref Stahl O. 1986, A\&A 164, 321
\ref Stahl O. 1987, A\&A 182, 229
\ref Tapia M., Roth M., Marraco H., Ruiz M.T. 1988, MNRAS 232, 661
\ref Thackeray A.D. 1950, MNRAS 110, 524
\ref Thackeray A.D. 1977, MNRAS 180, 95
\ref Viotti R., 1971, PASP 83, 170
\ref Viotti R., Altamore A., Barylak M., Cassatella A., Gilmozzi R.,
   Rossi C. 1984, in: Future of Ultraviolet Astronomy based on Six
   Years of IUE Research. NASA CP-2349, Washington, p.231
\ref Viotti R., Rossi L., Altamore A., Rossi C., Cassatella A. 1986,
   Luminous Stars and Associations in Galaxies. In:
   de Loore C.W.H., Willis A.J., Laskarides P. (eds.) Proc. IAU Symp.116,
   Reidel, Dordrecht, p. 249
\ref Viotti R., Cassatella A., Ponz D., Th\'e P.S. 1988, A\&A 190, 333
   (Paper II)
\ref Viotti R., Rossi L., Cassatella A., Altamore A., Baratta G.B.,
     1989, ApJS 71, 983
\ref Viotti R., Altamore A., Rossi C., Baratta G.B., Damineli Neto A.
     1992, High Resolution Spectroscopy with the VLT. In: Proc. ESO
     Workshop, ESO, Garching, in press
\ref Wang L., Mazzali P. 1992, Nature, 355, 58
\ref Whitelock P.A. 1992, South African Astronomical Observatory 1991 Report,
     Foundation for Research Development, Observatory, p.25
\ref Whitelock P.A., Carter P.S., Roberts G.,
     Whittet D.C.B., Baines D.W.T. 1983, MNRAS, 205, 577
\ref Williams P.M., van der Hucht K.A. 1992,
     Nonisotropic and variable outflows  from stars.
     In: L. Drissen, C. Leitherer, A. Nota (eds).
     ASP Conference Series, Vol.22, p.269
\ref Wolf B., Stahl O. 1982, A\&A 112, 111
\endref
\bye
%
\baselineskip=8pt
\font\pr=cmr7
\nopagenumbers
%
\noindent
{\pr
{\bf Table 2.} The optical spectrum of AG Car in June 1990
\halign{#
\hfill&#\hfill&#\hfill&#\hfill&#\hfill&#\hfill&#\hfill&#\hfill\cr
\noalign {\medskip}
\noalign {\hrule}
\noalign {\medskip}
Bar Wl & A/E~~  & Wid~ & Eqw~~ & Cont & ~Ion~~ &
Lab Wl & ~~Remarks \cr
\noalign {\smallskip}
{}~~~(1) & ~(2) & ~(3) & ~(4) & ~(5)& ~~(6) & ~~~(7) & ~~~~(8)\cr
\noalign {\smallskip}
\noalign {\hrule}
\noalign {\medskip}
434.18  &~E   & 1.2  & 0.456  & 186   & H~I & 434.047  &~~\cr
438.67  &~A  &0.6   &0.056  & 173&   He I  &438.793  & PCy  \cr
438.97  &~E  & 0.6  &0.034 &  171&   He I &   438.793 & ~\cr
439.44  &~A  &~     &    p  &~~    &    Fe III &  439.578 & ~\cr
439.68  &~E   &~    &    p  & ~    &    Fe III &  439.578  & ~\cr
441.80  &~A   &~    &    p  & ~    &  Fe III   &441.959 &  PCy \cr
442.10  &~E   &0.5  &0.018 & 162 &  Fe III &  441.959   & ~\cr
442.93  &~A   &~    &    p  &     ~&Fe III  & 443.095 &  PCy \cr
443.40  &~E   &0.5  &0.020 & 160  & Fe III &  443.095 & ~\cr
443.63 & ~A  & 0.4 &  0.014 & 168 &  He I  &   443.755   & ~\cr
444.82 & ~E  & 0.7 & 0.031 & 173 &  N II  &   444.703 & ~\cr
446.94 & ~A  & 0.6 &  0.069 & 178 &  He I  &   447.148 &  PCy \cr
447.30 & ~E  & 0.8 & 0.248 & 179 &  He I  &   447.148 & ~\cr
450.19  &~A  &~    &     p &~~    &Al III &  452.918& ~\cr
450.43  &~E   &1.0  &0.024 & 172  & Al III&   452.918&   red bl\cr
453.09  &~E   &0.9  &0.025 & 171  & Al III&   452.918& ~\cr
455.15  &~A   &0.6  & 0.044 & 173  & Si III&   455.265 &  PCy \cr
455.45  &~E   &0.6  &0.026 & 173  & Si III&   455.265   & ~\cr
456.69  &~A   &0.6  & 0.031 & 171  & Si III&   456.787&PCy, blue wing \cr
456.92  &~E   &0.5  &0.021 & 171  & Si III&   456.787    & ~\cr
457.40  &~A   &0.4  & 0.016 & 172  & Si III&   457.478&   PCy  \cr
457.69  &~E   &~    &    p  & ~    &Si III &  457.478   & ~\cr
460.05  &~A   &~    &    p  & ~    &   N II&     460.148&   PCy \cr
460.29  &~E   &0.5  &0.025 & 216  & N II  &   460.148   & ~\cr
460.59  &~A  &~     &    p  &  ~   &  N II &    460.715 &  PCy \cr
460.88  &~E   &0.7  &0.019 & 181  & N II  &   460.715   & ~\cr
461.24  &~A  &~     &    p  & ~    & N II  &   461.387 &  PCy \cr
461.47  &~E  &~     &0.020&  183 &  N II &    461.387   & ~\cr
462.06  &~A  &~     &  0.011 & 184 &  N II &    462.139 &  PCy  \cr
462.30  &~E  & 0.5  &0.018  & 185 &  N II &    462.139    & ~\cr
462.60  &~E  &~     &    p   & ~   &  C II &    462.571 &  bl \cr
462.89  &~A   &0.6  & 0.022  &186  & N II  &   463.054 &  PCy \cr
463.24  &~E   &0.5  &0.042  &188  & N II  &   463.054   & ~\cr
464.13  &~A   &0.3  & 0.025  &187  & N II  &   464.309 &   PCy \cr
464.43  &~E   &0.7  &0.016  &187  & N II  &   464.309   & ~\cr
464.84  &~A   &0.3  & 0.020  &187  & O II  &   464.914 &  PCy \cr
465.02  &~E   &~    &    p   &~~   &   O II&     464.914&   weak\cr
465.78  &~E   &1.1  &0.053  &187  &[Fe III]&  465.81 &   br \cr
468.62  &~E   &1.0  &0.032  &189  & He II  &  468.568 &  br \cr
470.10  &~E   &1.1  &0.027  &190  &[Fe III] & 470.15 & br, Al III 01.65\cr
471.10  &~A   &0.7  & 0.032  &190  & He I  & 471.314 &  PCy \cr
471.46  &~E   &1.0  &0.114  &190  & He I  &   471.314   & ~\cr
478.90  &~E   &0.8  &0.026  &197   &N II   & 478.813   & ~\cr
480.37  &~E   &1.0  &0.042  &200   &N II   &  480.327   & ~\cr
486.22  &~E   &1.9  &1.224  &204   &H I    &  486.133 &   \cr
624.22  &~E   &~    &    p   & ~    &   N II &    624.252& ~\cr
628.31  &~A   &2.3  & 0.102  & 129  & i.s.   &  628.3 &    i.s., db \cr
634.09  &~E   &~    &   ~~  & ~    & N II   &  634.067 &  bl \cr
634.62  &~E   &1.9  &0.120  &134   &N II    & 634.71&    Si II 4.71\cr
635.61  &~E   &~    &    p   & ~    &   N II &    635.70& ~\cr
637.05  &~E   &1.3  &0.035 & 134  & Si II  &  637.136&   br \cr
638.15  &~E   &~    &   p   & ~     &  N II  &   639.763& ~\cr
647.88  &~A   &~    &    p  &  ~    &N II    & 648.207 &  PCy\cr
648.32  &~E   &1.7  &0.119 &132   &N II    & 648.207& ~\cr
654.84  &~E   &~    &  ~~  &  ~~ & [N II]  &  654.81  &  bl\cr
656.39  &~E   &8.0  &2.820 &134   &H I     & 656.282      & ~\cr
658.38  &~E   &~    &    ~~ &  ~~ & [N II]  &  658.36 &   bl \cr
661.04  &~E   &1.1  &0.102 &134 &  N II    & 661.058& ~\cr
663.02  &~E   &~    &     p  & ~   & N II    & 663.05     & ~\cr
667.45  &~A   &0.8  &  0.097 & 135 & He I    & 667.815 &  PCy\cr
667.89  &~E   &1.9  &0.424 &132  & He I    & 667.815& ~\cr
705.92  &~A   &~    &     p  & ~  &     He I &    706.519&   PCy\cr
706.69  &~E   &2.5  &0.883 & 115 &  He I   &  706.519& ~\cr
722.40  &~A   &~    &   p    & ~   &  C II &    723.112 &  PCy \cr
722.95  &~E   &~    &    ~~  & ~  &  C II  &   723.112& bl \cr
723.61  &~E   &1.7  &0.094 &103  & C II   &  723.619 &  \cr
\noalign {\medskip}
\noalign {\hrule}}
}
%
\end